\documentstyle[10pt,epsfig,dp_delphititle,float,hangcaption,xspace,amssymb,
amsfonts,amsmath,amsthm,cite,graphicx]{dp_delphi}
\makeindex
\def\DpPaperGroup{PH-EP}
\def\DpPaperRef{2007-008}
\def\DpDate{28 February 2007}
\def\DpAuthors{DELPHI Collaboration\footnote{
Corresponding author: J. Timmermans, NIKHEF, P.O. Box 41882,
1009 DB Amsterdam, The Netherlands; e-mail: Jan.Timmermans@cern.ch}
}
\def\DpSubmit{(Accepted by Astropart. Phys.)}
\def\DpTitle{{ Study of multi-muon bundles in cosmic ray
showers detected with the DELPHI detector at LEP }}
\def\DpComment{}
\def\DpEMail{}

\newcommand{\delphi}{\textsc{DELPHI}\xspace}
\newcommand{\lep}{\textsc{LEP}\xspace}
\newcommand{\corsika}{\textsc{CORSIKA}\xspace}
\newcommand{\qgsjet}{\textsc{QGSJET}\xspace}

\newcommand{\eV}{\ifmmode \mathrm{eV} \else eV \fi}
\newcommand{\GeV}{\ifmmode \mathrm{GeV} \else GeV \fi}
\newcommand{\meters}{\ifmmode \mathrm{m} \else m \fi}
\begin{document}
\makeatletter
\makeatother

\begin{titlepage}
\pagenumbering{roman}

\CERNpreprint{\DpPaperGroup}{\DpPaperRef}   
\date{{\small\DpDate}}                      
\title{\DpTitle}                            
\address{\DpAuthors}                        

\begin{shortabs}                            
\noindent
The \delphi detector at \lep has been used to measure multi-muon
bundles originating from cosmic ray interactions with air. The cosmic
events were recorded in ``parasitic mode'' between individual $e^+e^-$
interactions and the total live time of this data taking is equivalent
to $1.6\cdot 10^6$ 
seconds. The DELPHI apparatus is located about 100 metres underground and
the 84 metres rock overburden imposes  a cut-off of about 52 GeV/c on
muon momenta. 
The data from the large volume Hadron Calorimeter allowed
the muon multiplicity of 54201 events to be 
reconstructed. 
The resulting muon
multiplicity distribution is compared with the prediction of the Monte Carlo
simulation based on CORSIKA/QGSJET01. The model fails to describe the
abundance of high multiplicity events. The impact of QGSJET internal
parameters on the results is also studied. 

\vspace{2cm}
\centering {\it This paper is dedicated to the memory of Heiner Herr}

\end{shortabs}

\vfill

\begin{center}
\DpSubmit \ \\          
\DpComment \ \\
\DpEMail \ \\
\end{center}


\vfill
\clearpage

\headsep 10.0pt

\addtolength{\textheight}{10mm}

\addtolength{\footskip}{-5mm}
\begingroup
%
\newcommand{\DpName}[2]{\hbox{#1$^{\ref{#2}}$},\hfill}
\newcommand{\DpNameTwo}[3]{\hbox{#1$^{\ref{#2},\ref{#3}}$},\hfill}
\newcommand{\DpNameThree}[4]{\hbox{#1$^{\ref{#2},\ref{#3},\ref{#4}}$},\hfill}
\newskip\Bigfill \Bigfill = 0pt plus 1000fill
\newcommand{\DpNameLast}[2]{\hbox{#1$^{\ref{#2}}$}\hspace{\Bigfill}}

\small
\noindent
\DpName{J.Abdallah}{LPNHE}
\DpName{P.Abreu}{LIP}
\DpName{W.Adam}{VIENNA}
\DpName{P.Adzic}{DEMOKRITOS}
\DpName{T.Albrecht}{KARLSRUHE}
\DpName{R.Alemany-Fernandez}{CERN}
\DpName{T.Allmendinger}{KARLSRUHE}
\DpName{P.P.Allport}{LIVERPOOL}
\DpName{U.Amaldi}{MILANO2}
\DpName{N.Amapane}{TORINO}
% ...

\normalsize
\endgroup
\newpage

\titlefoot{Department of Physics and Astronomy, Iowa State
     University, Ames IA 50011-3160, USA
    \label{AMES}}
\titlefoot{IIHE, ULB-VUB,
     Pleinlaan 2, B-1050 Brussels, Belgium
    \label{BRUSSELS}}
\titlefoot{Physics Laboratory, University of Athens, Solonos Str.
     104, GR-10680 Athens, Greece
    \label{ATHENS}}
\titlefoot{Department of Physics, University of Bergen,
     All\'egaten 55, NO-5007 Bergen, Norway
    \label{BERGEN}}
\titlefoot{Dipartimento di Fisica, Universit\`a di Bologna and INFN,
     Via Irnerio 46, IT-40126 Bologna, Italy
    \label{BOLOGNA}}
\titlefoot{Centro Brasileiro de Pesquisas F\'{\i}sicas, rua Xavier Sigaud 150,
     BR-22290 Rio de Janeiro, Brazil
    \label{BRASIL-CBPF}}
\titlefoot{Inst. de F\'{\i}sica, Univ. Estadual do Rio de Janeiro,
     rua S\~{a}o Francisco Xavier 524, Rio de Janeiro, Brazil
    \label{BRASIL-IFUERJ}}
\titlefoot{Coll\`ege de France, Lab. de Physique Corpusculaire, IN2P3-CNRS,
     FR-75231 Paris Cedex 05, France
    \label{CDF}}
\titlefoot{CERN, CH-1211 Geneva 23, Switzerland
    \label{CERN}}
\titlefoot{Institut de Recherches Subatomiques, IN2P3 - CNRS/ULP - BP20,
     FR-67037 Strasbourg Cedex, France
    \label{CRN}}
\titlefoot{Now at DESY-Zeuthen, Platanenallee 6, D-15735 Zeuthen, Germany
    \label{DESY}}
\titlefoot{Institute of Nuclear Physics, N.C.S.R. Demokritos,
     P.O. Box 60228, GR-15310 Athens, Greece
    \label{DEMOKRITOS}}
\titlefoot{FZU, Inst. of Phys. of the C.A.S. High Energy Physics Division,
     Na Slovance 2, CZ-182 21, Praha 8, Czech Republic
    \label{FZU}}
\titlefoot{Dipartimento di Fisica, Universit\`a di Genova and INFN,
     Via Dodecaneso 33, IT-16146 Genova, Italy
    \label{GENOVA}}
\titlefoot{Institut des Sciences Nucl\'eaires, IN2P3-CNRS, Universit\'e
     de Grenoble 1, FR-38026 Grenoble Cedex, France
    \label{GRENOBLE}}
\titlefoot{Helsinki Institute of Physics and Department of Physical Sciences,
     P.O. Box 64, FIN-00014 University of Helsinki, 
     \indent~~Finland
    \label{HELSINKI}}
\titlefoot{Joint Institute for Nuclear Research, Dubna, Head Post
     Office, P.O. Box 79, RU-101 000 Moscow, Russian Federation
    \label{JINR}}
\titlefoot{Institut f\"ur Experimentelle Kernphysik,
     Universit\"at Karlsruhe, Postfach 6980, DE-76128 Karlsruhe,
     Germany
    \label{KARLSRUHE}}
\titlefoot{Institute of Nuclear Physics PAN,Ul. Radzikowskiego 152,
     PL-31142 Krakow, Poland
    \label{KRAKOW1}}
\titlefoot{Faculty of Physics and Nuclear Techniques, University of Mining
     and Metallurgy, PL-30055 Krakow, Poland
    \label{KRAKOW2}}
\titlefoot{Universit\'e de Paris-Sud, Lab. de l'Acc\'el\'erateur
     Lin\'eaire, IN2P3-CNRS, B\^{a}t. 200, FR-91405 Orsay Cedex, France
    \label{LAL}}
\titlefoot{School of Physics and Chemistry, University of Lancaster,
     Lancaster LA1 4YB, UK
    \label{LANCASTER}}
\titlefoot{LIP, IST, FCUL - Av. Elias Garcia, 14-$1^{o}$,
     PT-1000 Lisboa Codex, Portugal
    \label{LIP}}
\titlefoot{Department of Physics, University of Liverpool, P.O.
     Box 147, Liverpool L69 3BX, UK
    \label{LIVERPOOL}}
\titlefoot{Dept. of Physics and Astronomy, Kelvin Building,
     University of Glasgow, Glasgow G12 8QQ
    \label{GLASGOW}}
\titlefoot{LPNHE, IN2P3-CNRS, Univ.~Paris VI et VII, Tour 33 (RdC),
     4 place Jussieu, FR-75252 Paris Cedex 05, France
    \label{LPNHE}}
\titlefoot{Department of Physics, University of Lund,
     S\"olvegatan 14, SE-223 63 Lund, Sweden
    \label{LUND}}
\titlefoot{Universit\'e Claude Bernard de Lyon, IPNL, IN2P3-CNRS,
     FR-69622 Villeurbanne Cedex, France
    \label{LYON}}
\titlefoot{Dipartimento di Fisica, Universit\`a di Milano and INFN-MILANO,
     Via Celoria 16, IT-20133 Milan, Italy
    \label{MILANO}}
\titlefoot{Dipartimento di Fisica, Univ. di Milano-Bicocca and
     INFN-MILANO, Piazza della Scienza 3, IT-20126 Milan, Italy
    \label{MILANO2}}
\titlefoot{IPNP of MFF, Charles Univ., Areal MFF,
     V Holesovickach 2, CZ-180 00, Praha 8, Czech Republic
    \label{NC}}
\titlefoot{NIKHEF, Postbus 41882, NL-1009 DB
     Amsterdam, The Netherlands
    \label{NIKHEF}}
\titlefoot{National Technical University, Physics Department,
     Zografou Campus, GR-15773 Athens, Greece
    \label{NTU-ATHENS}}
\titlefoot{Physics Department, University of Oslo, Blindern,
     NO-0316 Oslo, Norway
    \label{OSLO}}
\titlefoot{Dpto. Fisica, Univ. Oviedo, Avda. Calvo Sotelo
     s/n, ES-33007 Oviedo, Spain
    \label{OVIEDO}}
\titlefoot{Department of Physics, University of Oxford,
     Keble Road, Oxford OX1 3RH, UK
    \label{OXFORD}}
\titlefoot{Dipartimento di Fisica, Universit\`a di Padova and
     INFN, Via Marzolo 8, IT-35131 Padua, Italy
    \label{PADOVA}}
\titlefoot{Rutherford Appleton Laboratory, Chilton, Didcot
     OX11 OQX, UK
    \label{RAL}}
\titlefoot{Dipartimento di Fisica, Universit\`a di Roma II and
     INFN, Tor Vergata, IT-00173 Rome, Italy
    \label{ROMA2}}
\titlefoot{Dipartimento di Fisica, Universit\`a di Roma III and
     INFN, Via della Vasca Navale 84, IT-00146 Rome, Italy
    \label{ROMA3}}
\titlefoot{DAPNIA/Service de Physique des Particules,
     CEA-Saclay, FR-91191 Gif-sur-Yvette Cedex, France
    \label{SACLAY}}
\titlefoot{Instituto de Fisica de Cantabria (CSIC-UC), Avda.
     los Castros s/n, ES-39006 Santander, Spain
    \label{SANTANDER}}
\titlefoot{Inst. for High Energy Physics, Serpukov
     P.O. Box 35, Protvino, (Moscow Region), Russian Federation
    \label{SERPUKHOV}}
\titlefoot{J. Stefan Institute, Jamova 39, SI-1000 Ljubljana, Slovenia
    \label{SLOVENIJA1}}
\titlefoot{Laboratory for Astroparticle Physics,
     University of Nova Gorica, Kostanjeviska 16a, SI-5000 Nova Gorica, Slovenia
    \label{SLOVENIJA2}}
\titlefoot{Department of Physics, University of Ljubljana,
     SI-1000 Ljubljana, Slovenia
    \label{SLOVENIJA3}}
\titlefoot{Fysikum, Stockholm University,
     Box 6730, SE-113 85 Stockholm, Sweden
    \label{STOCKHOLM}}
\titlefoot{Dipartimento di Fisica Sperimentale, Universit\`a di
     Torino and INFN, Via P. Giuria 1, IT-10125 Turin, Italy
    \label{TORINO}}
\titlefoot{INFN,Sezione di Torino and Dipartimento di Fisica Teorica,
     Universit\`a di Torino, Via Giuria 1,
     IT-10125 Turin, Italy
    \label{TORINOTH}}
\titlefoot{Dipartimento di Fisica, Universit\`a di Trieste and
     INFN, Via A. Valerio 2, IT-34127 Trieste, Italy
    \label{TRIESTE}}
\titlefoot{Istituto di Fisica, Universit\`a di Udine and INFN,
     IT-33100 Udine, Italy
    \label{UDINE}}
\titlefoot{Univ. Federal do Rio de Janeiro, C.P. 68528
     Cidade Univ., Ilha do Fund\~ao
     BR-21945-970 Rio de Janeiro, Brazil
    \label{UFRJ}}
\titlefoot{Department of Radiation Sciences, University of
     Uppsala, P.O. Box 535, SE-751 21 Uppsala, Sweden
    \label{UPPSALA}}
\titlefoot{IFIC, Valencia-CSIC, and D.F.A.M.N., U. de Valencia,
     Avda. Dr. Moliner 50, ES-46100 Burjassot (Valencia), Spain
    \label{VALENCIA}}
\titlefoot{Institut f\"ur Hochenergiephysik, \"Osterr. Akad.
     d. Wissensch., Nikolsdorfergasse 18, AT-1050 Vienna, Austria
    \label{VIENNA}}
\titlefoot{Inst. Nuclear Studies and University of Warsaw, Ul.
     Hoza 69, PL-00681 Warsaw, Poland
    \label{WARSZAWA}}
\titlefoot{Now at University of Warwick, Coventry CV4 7AL, UK
    \label{WARWICK}}
\titlefoot{Fachbereich Physik, University of Wuppertal, Postfach
     100 127, DE-42097 Wuppertal, Germany \\
\noindent
{$^\dagger$~deceased}
    \label{WUPPERTAL}}

\addtolength{\textheight}{-10mm}
\addtolength{\footskip}{5mm}
\clearpage

\headsep 30.0pt
\end{titlepage}

%
\pagenumbering{arabic}                              
\setcounter{footnote}{0}                            %
\large

\section{Introduction}

\begin{figure}[e]\centering  
\epsfig{file=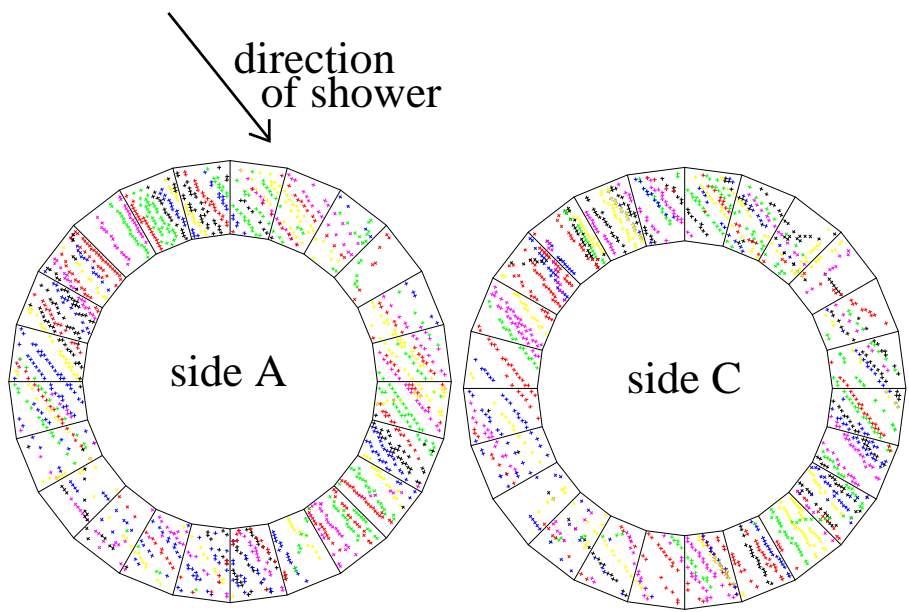,height=8cm}
\caption{High multiplicity cosmic event as seen by hadron calorimeter.
The number of reconstructed tracks was $127$.}
\label{niceevt}
\end{figure}

The \delphi (DEtector with Lepton Photon and Hadron Identification) at
CERN \lep (Large  Electron Positron collider)
measured cosmic  muons regularly in order to align and calibrate
various subdetectors. A  major upgrade of the 
\delphi hadron calorimeter was completed in 1997. As a
result the calorimeter granularity increased substantially and
spectacular events like the one shown in Fig.~\ref{niceevt} were
registered. The trigger studies performed during 1998 have shown that
\delphi can register cosmic events during regular data
taking. Whenever there was no triggered $e^+e^-$ interaction, the detector
stayed active to record possible cosmic events. In this regime we were
able to collect data throughout the years 1999 and 2000. 

The experimental hall of \delphi was located 100 metres underground
and the overburden imposed a cut-off of 52 GeV/c on the momenta of vertical
muons. This,
depending on the particular interaction model, corresponds to a lower limit
of primary particle energies of about $10^{14}~$eV. The upper limit of
primary energy, less than $10^{18}$ eV, follows from the total
measurement time of $1.6\cdot 10^6$ seconds. Although this live time is small
compared to standard cosmic ray experiments, the granularity of the 
detector and the momentum cut-off make the data interesting. The high
energy muons originate from meson decays and other processes which
take place in the upper atmosphere. They carry information about the
first stages of the shower  development. Consequently, these data
reflect different aspects of the shower than those recorded by
experiments on the ground, where the vast majority of detected muons
originates from pion decays at low energies. 

Reconstruction of cosmic ray interactions at very high energies relies
heavily on Monte Carlo (MC) simulations. Hence the interpretation of
measured data is dependent on the 
models of shower propagation, including simulations of high energy
hadron collisions, hadron decays and further development of the
electromagnetic and hadronic components. 
While the particle decays and the shower propagation are well
described, the most important source of uncertainties originates from
models describing the high energy interactions of hadrons at the
beginning of shower development. The interaction models such as
NEXUS\cite{nexus}, QGSJET\cite{qgsjet} or SIBYLL\cite{sibyll} are tuned
to available  accelerator data at lower energies than those discussed
in this paper. The collider experiments are more suited to study
phenomena at larger transverse  momenta. Thus our data, which can
reveal features of particle interactions in the very forward region, are
in this sense complementary. 

The muon component of cosmic ray showers has been studied with large
ground arrays (e.g. \cite{kascademuon,casa-mia}) or at large depths
corresponding to a momentum cut-off above  
1 TeV (e.g. \cite{frejus,KGFmulti,macro}). The data at intermediate
depths underground are scarce and the experiments detecting muons with
a momentum cut-off around 100 GeV/c (e.g. \cite {baksan}) use less precise
detectors than the \lep experiments. Besides \delphi, similar studies of
cosmic rays were performed at ALEPH \cite{eggert} and L3+C
\cite{l3c}. 
Detailed model tests \cite{Testmodely} show that QGSJET describes best
the various
correlations between hadronic, electromagnetic and muon
 components of atmospheric showers in the case of ground experiments.
 Data registered by underground experiments reflect different shower
 properties.
   The aim of this work is  to test the
 interaction model, which  sufficiently well describes            the ground
 measurements,  using 
 multi-muon data  detected underground. 

The detector and its overburden are described in
Section~\ref{hware}. The conditions of event registration are
mentioned in Section~\ref{trig} and the procedure of event
reconstruction is described in Section~\ref{evrec}. The chain of
programs used to simulate showers is described in
Section~\ref{sim}. The results obtained are given together with
predictions of hadronic interaction models 
 in Section~\ref{res} and they are
discussed in the final Section~\ref{disc}.

\section{Detector and its location}
\label{hware}

\begin{figure}[h!]
\centering
\epsfig{file=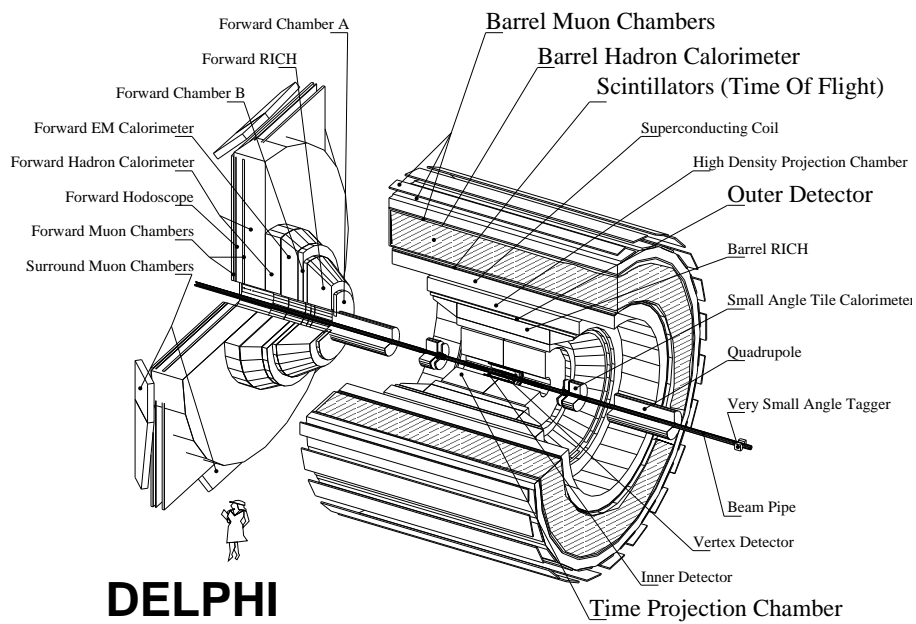,width=12cm}
\caption{The layout of the \delphi detector; the hatched area
represents the hadron calorimeter. Subdetectors used in this work are
marked by larger letters.}
\label{delphidet}
\end{figure}

\delphi was a classical collider experiment with numerous subdetectors
and a solenoidal magnetic field. A detailed description of the apparatus
can be found in \cite{delphidet}. Only a few subdetectors were used
for the cosmic muon detection, namely: Time Projection Chamber (TPC),
Time Of Flight scintillation detector (TOF), Outer Detector (OD),
Barrel part of HAdron calorimeter (HAB) and Barrel MUon chambers
(MUB). All these parts were located in the barrel part of the detector
(Fig.~\ref{delphidet}). TOF served to trigger cosmic events.

The HAB detector was a sampling calorimeter and it contained $12000$
limited streamer tubes.  
The iron of the magnet yoke served as an absorber. It consisted of 20
slabs 5 cm thick. Streamer tubes were inserted into the 2 cm wide gaps
between individual iron plates. The gas mixture inside the tubes was
composed of Ar($10\%$), CO$_2$($60\%$) and iso-butane($30\%$).  
HAB with its large volume served as the backbone of muon
detection. The detection area of HAB was $75~{\meters}^2$ in
the horizontal plane.
Each tube in the  barrel part of the hadron calorimeter had an effective
length of 3.6 m and its cross-section was 1~$\times$~8~cm$^2$. 
 All the tubes were parallel to the beam
pipe. During the upgrade of the hadron calorimeter in the years 1995 - 1997
each tube was equipped with read-out of its cathode, which consisted
of resistive varnish of the whole tube interior \cite{delphidet2}. The
smallest sensitive cell before the upgrade was about $20 \times 30
\times 35~\mathrm{cm}^3$ in ($\theta$, $\phi$, $R$) standard DELPHI
coordinate system\footnote{as defined e.g. in \cite{delphidet} - $R$
  radius, $\phi$ azimuth angle in plane perpendicular to the beam pipe
  and $\theta$ polar angle ($=0$ along beam)} and the cells were
organised in towers 
pointing to the centre of the detector. After the upgrade the cell
size of the cathode readout in the barrel became $360 \times 8
\times 7~\mathrm{cm}^3$ \cite{aj1,aj2}. Consequently  the 
granularity in the plane perpendicular to the beams increased about
$14$ times. Due to 
technical limitations it was possible to read out signals only on the
two outer front-ends of the barrel. The charge deposited on the
cathode was integrated for 350 $\mu$s and accepted or rejected by
a discriminator. Thus in this system of cathode read-out, the signals
from individual tubes were either {\em yes} or {\em no} and the
reconstructed tracks are in fact only projections of the muon trajectories
onto the plane perpendicular to the \lep beams, separately for each half of
HAB.  

The TPC was able to measure the full direction of muon tracks. Due to
its relatively small volume it contained only a small fraction of the muons
passing through DELPHI (TPC had $10$ times smaller detection area
compared to HAB). During the standard recording of $e^+e^-$ collisions, the
drift time in the TPC is measured from $t_0$ which is given by the instant
of beam cross-over (BCO) inside \delphi.  In the case of cosmic
events t$_0$ was the average arrival time  of tracks to the OD.

In extreme cases $50\%$ or more of the tubes in one or both sides of
HAB were hit. This led to saturated events where counting of individual
muons was not possible anymore.
 However, the cosmic origin of these events is guaranteed, because in
 this case vacant tubes appear in parallel lines which 
 follow the direction of the muon bundle and they 
 cannot be caused by  any noise in HAB.
 Moreover, in a few such events the lower bound on
 the number of muons could be roughly assessed from MUB.

\begin{figure}[h!]
\centering
\epsfig{file=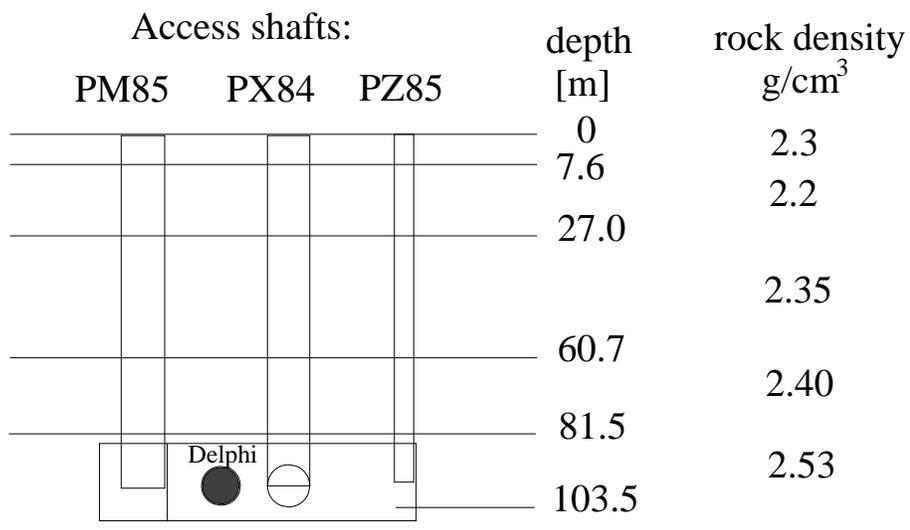,height=7cm}
\caption{Schematic picture of rock overburden above \delphi detector.}
\label{geom}
\end{figure}

The apparatus was situated about $100~{\meters}$ underground. The
surface altitude was $428~{\meters}$ above the sea level. The
composition of the rock above the \delphi experiment is known from
a geological survey performed for civil engineering purposes. The
simplified picture of the overburden structure could be approximated
by $5$ major geological layers with different mass densities. The density
of the rock in the vertical direction varies between $2.2~\mathrm{g}/
\mathrm{cm}^{3}$ and $2.5~\mathrm{g}/ \mathrm{cm}^{3}$ depending on
the layer. The total vertical depth of DELPHI location is about
$19640~\mathrm{g}/ \mathrm{cm}^{2}$. The resulting energy cut-off for
vertical cosmic muons is 
$\sim 52~{\GeV}$. The detector was located in a large experimental cavern
equipped with three access shafts shown in Fig.~\ref{geom}. This
scheme of the experimental area and the overburden was used in simulations.

\section{Trigger}    
\label{trig}

The trigger of cosmic events was entirely based on TOF. This detector
consisted of a single layer of plastic scintillation counters. Each
one was read out by two photomultipliers. The scintillator planks
covered the internal side of HAB. Initial attempts to trigger on single muons
led to a high trigger rate. Therefore in 1999 the trigger was set up to
demand at least $3$ active detector sectors to accept an event. It ran
in so-called ``parasitic mode'', i.e. whenever there was no triggered
$e^+e^-$ interaction, the trigger stayed sensitive to cosmic events
for $4.1~\mathrm{\mu s}$ after each beam crossing. This 
short detection window was optimised for $e^+e^-$ interactions.

The  beam crossing frequency depended on the number of $e^+$($e^-$)
bunches in the collider. During the running mode with $4$ bunches in
the machine, the beam 
crossing period was $\sim 22.2~\mathrm{\mu s}$, while in the $8$ bunch
mode the period decreased to 
$\sim 11.1~\mathrm{\mu s}$. Consequently, the detector was sensitive
to cosmic events for $18\%$ of  
the total data taking time in  $4$ bunch mode and for $37\%$ in $8$
bunch mode. Dedicated cosmic runs (without the beams in the
collider) have been performed mainly at the beginning of each
year. Although there were no $e^+e^-$ collisions, BCO signals were
issued to mimic the $8$ bunch mode.  

\begin{figure}[h!]
\centering
\epsfig{file=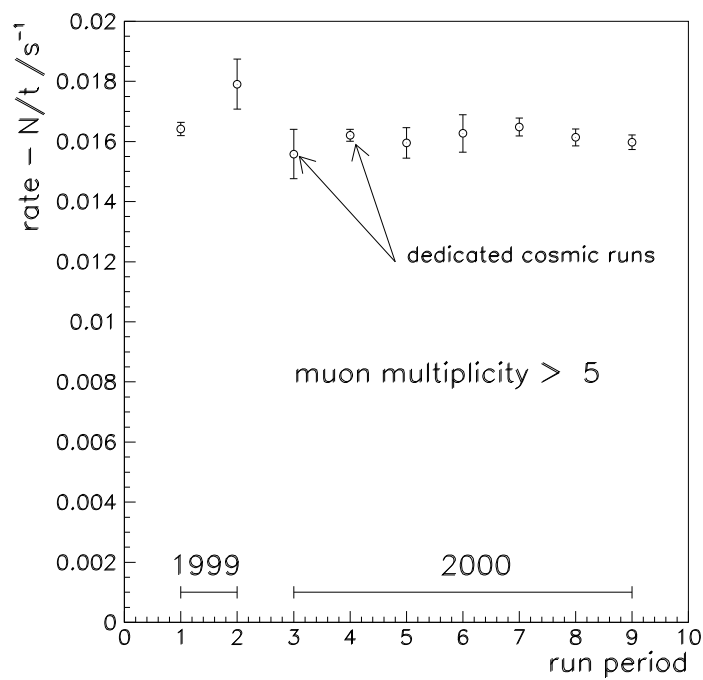,height=10cm}
\caption{Event rates ($N_{\mu} > 5$) for different run periods.}
\label{pictri14}
\end{figure}

In an ideal case, two muons passing TOF would be sufficient to
activate the trigger. 
In reality the TOF detection efficiency in $Z^0 \rightarrow \mu^+
\mu^-$ events was 
 84$\%$.
 However, with increasing muon multiplicity the
 TOF trigger efficiency quickly approaches almost $100\%$.
Already for  muon multiplicity $N_\mu = 5$ the TOF efficiency is
99$\%$, for lower multiplicities $N_\mu = 3(4)$ the corresponding
efficiencies are 94(97)$\%$.
 It was found in \cite{diser} that with $5$ or more muons the
  trigger stability is  assured.
 Fig.~\ref{pictri14} plots the rate of events
with muon multiplicity higher than $5$ in different run periods. The
event rates are consistent within   statistical errors and there is no
difference between the runs with and without beams in \lep.
In total, taking into account various bunch schemes and the
$4.1~\mathrm{\mu s}$ detection window, the accumulated effective live
time is $T_{eff} = 1.6\cdot10^6~\mathrm{s}\ (= 18.5\ \mathrm{days}$). 

\section{Event reconstruction}
\label{evrec}

The tracks of cosmic muons were reconstructed from hadron calorimeter
data by the ECTANA 
 program \cite{ECTANA}, which
 scans signals in the HAB modules  and finds
 track patterns  of hit 
 streamer tubes. This package has the advantage  that it was developed not
 only for studies  of   $e^+e^-$ collisions,  i.e. tracks coming from
 the interaction point in the centre of the 
 detector, but it has the option for cosmic events as well.
 When running in cosmic mode it allows tracks
 originating  anywhere  in the calorimeter to be reconstructed without
 an explicit cut 
 on the track  impact parameter. 
 The search for active streamer tubes starts from the  outer planes
 of a given  module and continues inwards. A group of at least $4$
 aligned  hits is taken as a track element. The track element is also
 required to have a reasonable density of hits, at least $30\%$ of tubes along
 its length  have  to be active.
 All possible hypotheses starting from a certain hit found during the
 scan are analysed, and the positions of hits are fitted by a straight line.
 The best fit in terms of the number of hits and $\chi^2$ is stored.
 Before accepting the track, its similarity with other hypotheses was
 checked to avoid double counting.

 The length of the reconstructed
track was required to be larger than $50~\mathrm{cm}$. It was possible
to fit radii of curvature of the bent tracks, however, there were only
a few such tracks and their radii were quite large. Therefore
the coordinates of active tubes
were fitted only by straight lines  in the 
final analysis. 
The matching between track elements from different calorimeter sectors
was performed.
The number of  reconstructed  tracks  was
considered as the reconstructed multiplicity of an event. The
performance and functionality of the ECTANA program were checked with MC
studies that compared
parameters of  reconstructed and injected events. However, no MC
tuning of the reconstruction software was needed. 

\begin{figure}[h!]
\centerline{\epsfig{file=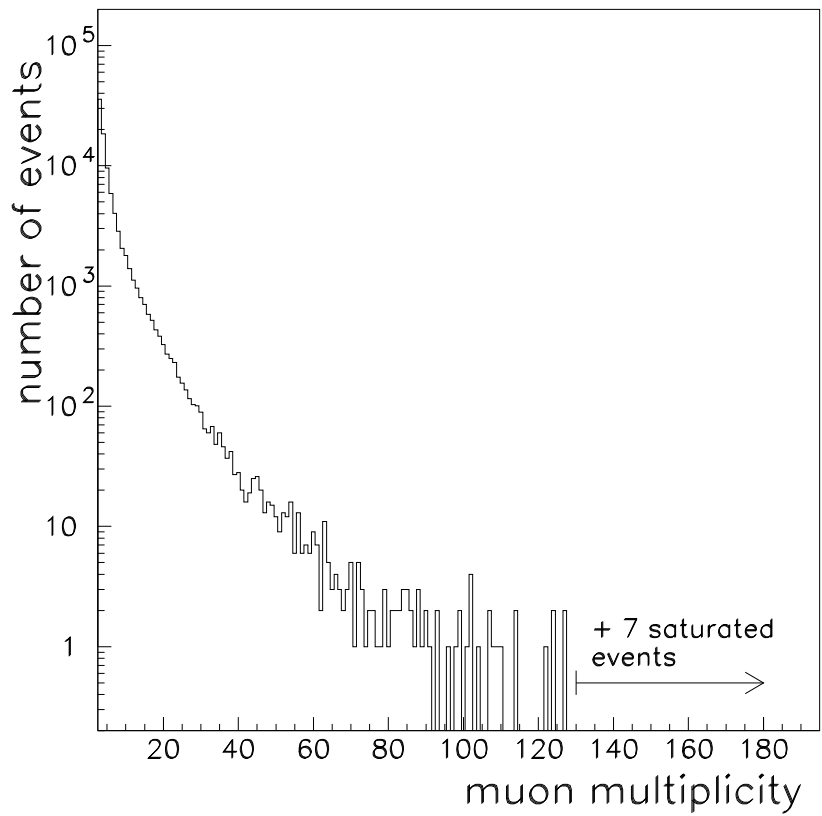,height=12cm,width=12cm}}
\caption{Differential muon multiplicity distribution.}
\label{mul}
\end{figure}

 The analysed data sample consists of $54201$ events with muon
multiplicities bigger 
than $3$. They were registered during the years $1999$ and $2000$.
The number of events with multiplicity above a given value is
given in Tab.~\ref{stat}  and the differential multiplicity distribution
is shown in Fig.~\ref{mul}.  

\begin{figure}[h!]\centering
 \epsfig{file=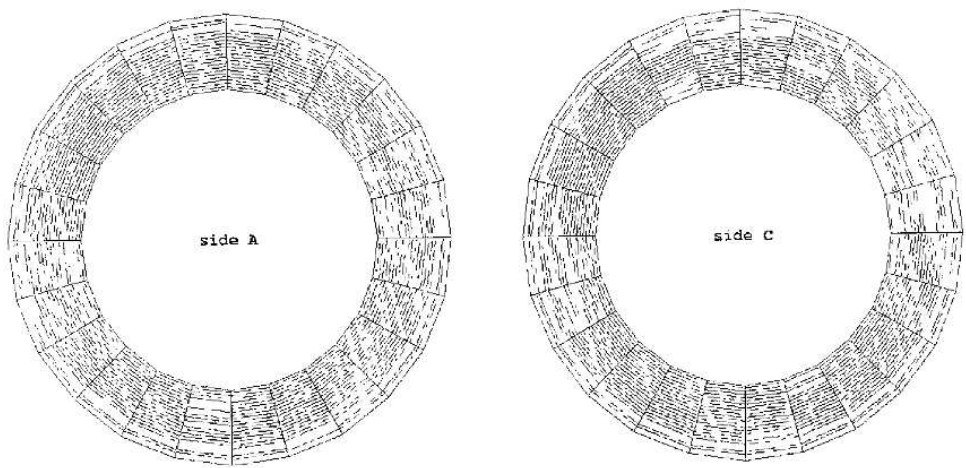,height=7cm}
\caption{A saturated event in the hadron calorimeter. Vacant
tubes show voids in the muon bundle. }
\label{satevt}
\end{figure}

\begin{figure}[h!]\centering
 \epsfig{file=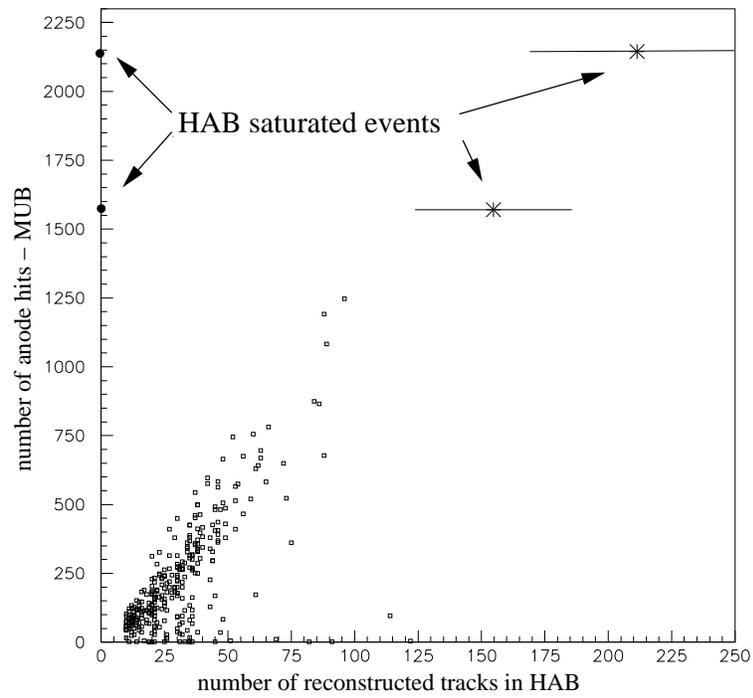,height=9.1cm}
\caption{Multiplicity reconstruction of saturated events from MUB data.}
\label{MUBsat}
\end{figure}

Altogether there were only 7 saturated events like the one depicted in
Fig.~\ref{satevt} where more then 50\% of the tubes were hit. 
In the case of saturated events vacant tubes make parallel line
patterns which cannot 
result from a glitch of the electronics.
The saturated events
are expected to have multiplicity higher than the highest multiplicity
reconstructed from unsaturated events  
($N_{\mu}>  127$).
Moreover, in two
of these events we were able to assess the lower limit of the multiplicity
 from the proportionality
 between the  number of MUB anode hits and reconstructed muon multiplicity from HAB
(Fig.~\ref{MUBsat}). However, this procedure was not possible in all
events. The MUB time window is only 5.9  $\mu$s after BCO and the events
coming at the end of trigger time window 4.1 $\mu$s after BCO are not
registered properly as the necessary drift time is 2.5 $\mu$s.

\begin{figure}[h!]\centering
 \epsfig{file=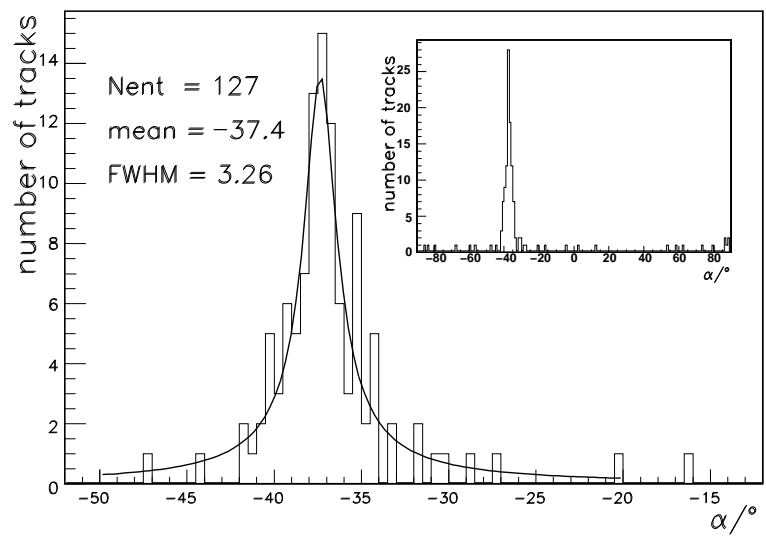,width=10cm}
\caption{Projected angle distribution in a high multiplicity event.}
\label{collinear}
\end{figure}

 In general, the muon tracks
inside bundles are almost parallel as demonstrated in
Fig.~\ref{collinear}. In this picture we plot the angle $\alpha$
between the vertical direction and the track 
projection onto the plane perpendicular to the \lep beams.
 The track collinearity helped to find high 
multiplicity events originating from muon interactions close to the
detector. The manual scanning was done on all events with $N_{\mu} >
30$. Altogether we have rejected 14 events with diverse
directions of tracks. 
 They correspond to $1.3 \%$
of the $1065$ scanned events. The parallelism of reconstructed tracks was
checked also by the cut that requires more than $50\%$ of
reconstructed tracks to be aligned within $5^\circ$ of the mean
value of all track angles in the event. This cut rejected the same
events as the scanning procedure. 

\begin{figure}[hbt]
\centering
\includegraphics[height=9cm]{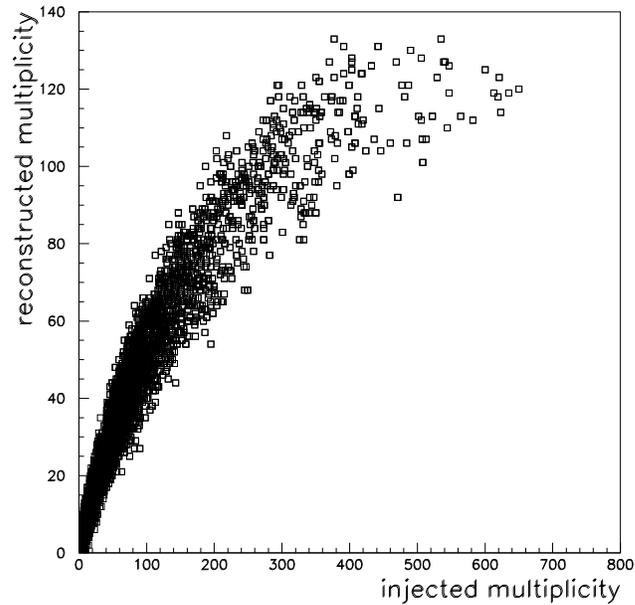}
 \caption{The correlation 
 between injected and reconstructed numbers of
   muons in MC simulation.}
\label{proportionality}
\end{figure}

As already mentioned above, the cathode read-out could not detect how
many muons hit one single tube. Therefore at higher multiplicities
muons start to shadow each other and the reconstructed multiplicity is
in fact a lower limit of the real event multiplicity. However, even the
highest reconstructed multiplicities around 120 are still strongly correlated
with the initial multiplicity as can be seen from
Fig.~\ref{proportionality}, where the reconstructed multiplicity in MC
data is plotted as a function of the number of muons injected into
HAB.

Unlike  the hadron calorimeter, the TPC gives full spatial information on
traversing muons. The drawback is its relatively small size. The track
reconstruction from the TPC was possible with standard \delphi software
tools with the provision for start of the drift time (see
Section~\ref{hware}). 
Due to the disproportion of TPC and HAB sizes, the
respective multiplicities do not correlate well. However, 
we were able
to reconstruct the muon bundle directions from the TPC and to compare the 
multiplicities from the TPC with MC predictions \cite{diser}.

\section{Simulation}
\label{sim}

To simulate the response of \delphi to cosmic-ray induced showers, we
have set up a chain of simulation programs. The high energy
interactions were modelled by the QGSJET01 \cite{qgsjet} program
implemented within the \corsika \cite{CORSIKA} package\footnote{First
  analyses with \qgsjet model were 
   performed with CORSIKA { \it ver. 6.014} from March 2002. Later
   studies of
  \qgsjet with modified parameters used CORSIKA { \it ver. 6.031} from
 February 2004. It was checked that the results of the two simulations were
 independent of the CORSIKA version.} . 
The rock above the \delphi
detector and the shape of the experimental cavern as well as the basic structures such as concrete walls and the three access shafts were represented according to Fig.~\ref{geom} and simulated by GEANT3\cite{GEANT}. Full simulation of the detector response was provided by the DELSIM \cite{delsim} simulation package.
  
As the chemical composition of cosmic rays is not well known, we have
used only two limiting cases of hadron primary particles - protons and
iron nuclei. Data sets were generated for both types of primary
particles in $12$ energy intervals $10^{12}$~-~$3$$\cdot$$10^{12}$~$\eV$,
$3$$\cdot$$10^{12}$~-~$10^{13}$~$\eV$, etc. up to
$3$$\cdot$$10^{17}$~-~$10^{18}$~$\eV$. The lowest energy interval barely
contributes  due to the muon energy cut-off of $52~{\GeV}$ and the
condition $N_{\mu} > 3$. Also the highest energy interval contributes
very little, if at all, because of the relatively short observation
time. The lower energy limit depends on the interaction model and
on the thickness of the overburden while the
upper limit is given by the flux value used for normalisation and the
observation time.  As these
two limits are not given reliably we have used  a wider energy
range for the simulations.

All CORSIKA simulations  were done without ``thinning''.
At high energies ($E>10^{17}$~$\eV$) the  thinning option  speeds up simulations
of showers with billions of secondary particles by discarding a defined
fraction of the secondaries and by ascribing the remaining particles certain 
weights. However, this option  might introduce additional systematic
errors. For this reason full event simulation was used  in the
analysis.

The data samples were generated according to an energy dependence $\sim
E^{-\gamma}$ using the spectral index $\gamma=1$
in order to obtain sufficient representation of events at the upper part of
the energy spectrum. Events were then re-weighted according to one of the
assumed energy spectra (see below).

\begin{figure}[hbt]
\begin{center}
\epsfig{file=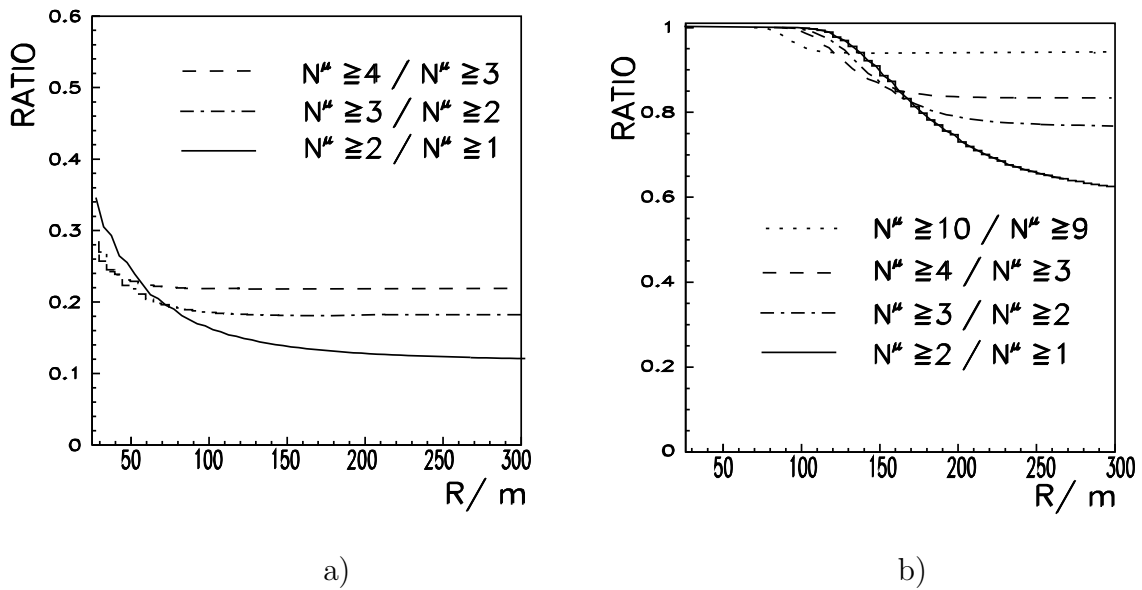,width=15cm}\\
\end{center}
\hspace{4.5cm} a) \hspace{7.0cm} b)
\begin{center}
\caption{Ratio of two adjacent 
 bins (see legends inside the plots)
 of integral multiplicity distribution 
  as a function of the parameter $R$.
  The plots correspond to iron induced vertical showers at a primary
  energy of $10^{14}~{\eV}$ (a)  
  and $10^{17}~{\eV}$ (b).} 
\label{ratios}
\end{center}
\end{figure}

Shower centres were smeared uniformly over a circular area with radius
$R=200~{\meters}$ around the \delphi detector.
 This radius value was chosen as optimal because smaller
$R$ values led to an increased fraction of lost events with small muon
multiplicities while  larger radii would imply the necessity of using 
large data samples to produce enough events with high muon
multiplicities. This is demonstrated in Fig.~\ref{ratios} which shows
the stability of the simulated multiplicity distribution as a function of
$R$.  For each radius the  ratio of
occupancies in two  adjacent
bins in the final integral multiplicity\footnote{Defined
   in Section~\ref{res}.} distribution is plotted.
 With increasing values of $R$, the simulated multiplicity distribution
 stabilises.  At  $R=200~\meters$  the
 stability is reached at all simulated energies.
 Furthermore, the radius  of $200~{\meters}$ ensures that the fraction of lost
 events at the lowest multiplicity $N_\mu = 4$ is smaller than 0.5\%. 

During the smearing of showers with  $E < 10^{16}~{\eV}$ each shower
was used $10$ times. For higher energies the number of moves is
$100$. 
  Taking $100$ moves at energy $>10^{16}~{\eV}$, 
  one CORSIKA generated shower contributed to the simulated spectrum at
  $N_{\mu} > 45$ on average  only once.
 Since the events with $N_{\mu} > 45$ are dominated by
  primary energies higher than  $10^{16}~{\eV}$, the relatively high
  number of moves is, in fact,  chosen optimally.
 
The generated data set at $N_{\mu} > 45$ (which corresponds roughly
to  $E  \sim 10^{16}~{\eV}$) was about 20 times larger than the real data
sample. At lower multiplicities (i.e. lower energies)  the samples
were about equal. 
 The  stability of the results was also checked by  doubling the size of
 the MC data sets. 

\begin{figure}[h!]
\centering
 \includegraphics[height=9cm]{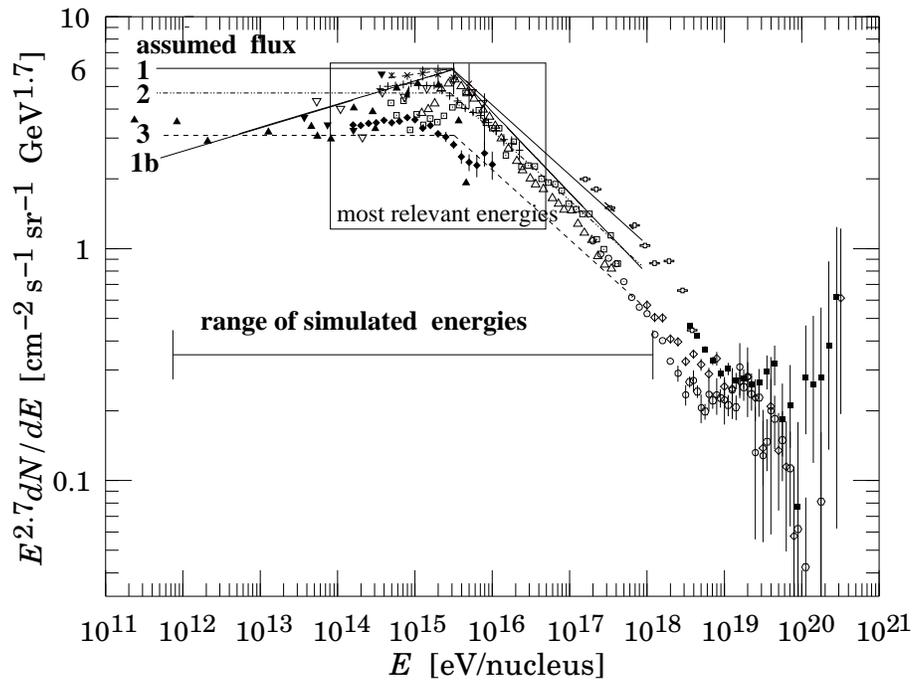}
\bf \caption{ \rm Assumed fluxes compared to various measurements. The
 picture is taken from \cite{pdg} and modified. The squares close to line {\it 1} correspond to
 results of Haverah Park taken from \cite{have}. The data points were added using the macro available at
 http://astroparticle.uchicago.edu/announce.html. Fluxes are multiplied by  $E^{2.7}$.}
 \label{asspec}
\end{figure}

The normalisation of the simulated multiplicity distributions depends
necessarily on the assumed energy spectrum of primary particles. Four
spectra corresponding to different lines in Fig.~\ref{asspec} were
assumed. Lines {\it 1, 2} and {\it 3} all represent power law indices
$\gamma = 2.7$ below the knee
($E_{\mathrm{knee}}=3\cdot10^{15}~{\eV}$) and $ \gamma = 3.0$ above
the knee, thus they have the same shape of energy dependence  and they differ
by the total flux only. 
 Assumption {\it 1b} is defined by exponents $\gamma = 2.6$
below  and $ \gamma = 3.05$ above the knee. These spectral indices
were used for tests of QGSJET01 with changed internal parameters. 

The most notable contributions to the systematic errors
are our imperfect knowledge of the overburden and due to a hardware effect
which in certain situations caused cross-talk of the cathode read-out and
appeared as a wider muon track
that can shadow more muons than the normal track.
The effect of inaccurate knowledge of the overburden was taken into
account by changing the rock density by $\pm 
5\%$ in all geological layers.  Changes of multiplicity
distribution  induced by this density variations stay within 5\%.
 The cross-talk has been studied in detail in $Z^0 \rightarrow \mu^+\mu^-$
 interactions. Based on this it was incorporated into the MC. The
 systematic error induced by this effect was checked  
in MC  by using two options: one with full cross-talk 
simulation taken into account and another with this simulation switched off.
It was found that the impact of cross-talk on the final
multiplicity distribution is less than  $ 5\%$ of the number
 of events at high multiplicities. The upper bound of the possible
 live time error was estimated using the knowledge of the  DELPHI dead
 time and it is  about $2\%$. Due to the DELPHI magnetic field,
 another  effect which might induce systematic error is the possible
 track  matching inefficiency in the upper and  the lower part of HAB
 for low energy muons. Assuming only straight lines in track
 reconstruction we could double count curved tracks.  The effect was
 studied using the  
 option of the ECTANA package that enabled to search also for curved
 tracks. It was found that the maximal impact on the final
 multiplicity   distribution decreases with increasing multiplicity
 and it is about  $8\%$ for  multiplicities  below $15$,  $4\%$ at 
 integrated multiplicities  larger than $20$,  $3\%$  for
 multiplicities larger than $45$ and $2\%$  for
 multiplicities larger than $70$. 

 The overall systematic error
 is $\lesssim  8\%$ at high multiplicities  ($N_\mu \geq 45$) which is
 below  the  statistical  uncertainty. 
 More detailed discussion of the whole
 simulation is provided in \cite{diser}. 

\section{Results}
\label{res}

\subsection{Directions of muon bundles}
\begin{figure}[!h]\centering
 \epsfig{file=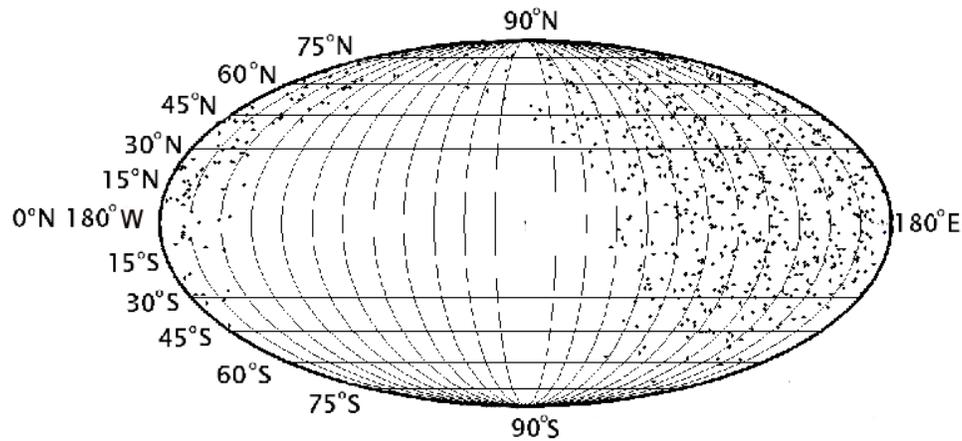,height=10.cm}
\bf \caption{ \rm Galactic coordinates of events with more than $15$
  tracks in HAB and 
 more than $3$ reconstructed tracks in TPC.}
\label{skyplot}
\end{figure}

The most straightforward and MC-independent results are those
concerning the directions of muon bundles. As explained already above,
it was possible to reconstruct the full spatial direction of the
tracks only from TPC data.
 As the TPC reconstruction depends on the mean arrival time to OD,
 we have selected higher multiplicity events with more than 15 muons in
 HAB and at least 4 reconstructed tracks in TPC. 
 This cut corresponds to primary energies of about $\sim 10^{15}~{\eV}$.
 The sky plot of event directions in galactic
coordinates is shown in Fig.~\ref{skyplot}. The event direction is
given as a mean direction of individual muons and the pointing
precision is a few degrees due to multiple
scattering in the overburden, detector precision and unknown core
position of the shower. 
 There is no apparent clustering of events.  

\begin{figure}[!h]\centering
 \epsfig{file=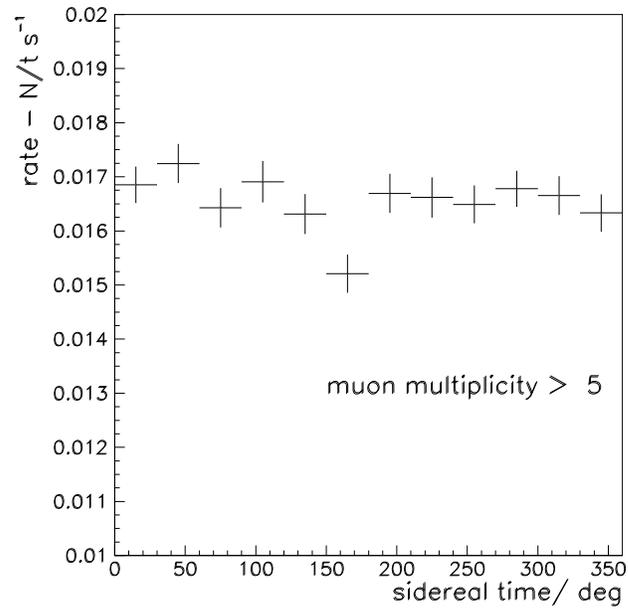,height=9.cm}
\bf \caption{ \rm The event rate versus the sidereal time expressed in
degrees. Events with more than $5$
 reconstructed muons are taken into account.}
\label{sidereal}
\end{figure}

The absence of point sources is demonstrated also by the dependence of the
event rate on sidereal time. Fig.~\ref{sidereal} shows no significant
modulation of the rate during the sidereal day. The small dip
disappears at higher  multiplicities. 

The lack of point like anisotropies in the data justifies the assumption of
uniform distribution of cosmic ray directions which is used in
MC simulations.

\subsection{Muon multiplicities}
\begin{figure}[!h]\centering
  \epsfig{file=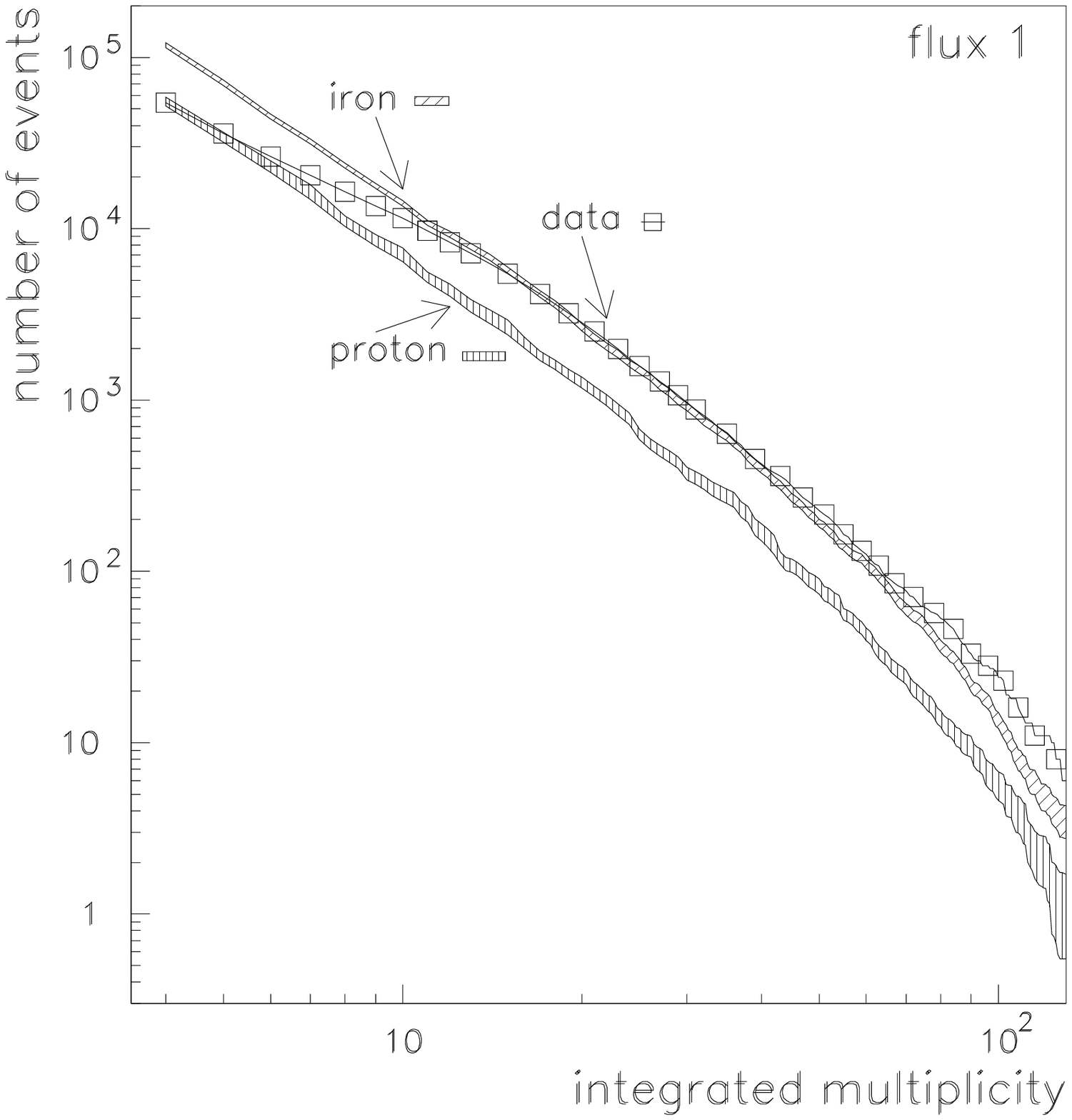,width=7.3cm, height=7.3cm}\hspace{1.cm}
  \epsfig{file=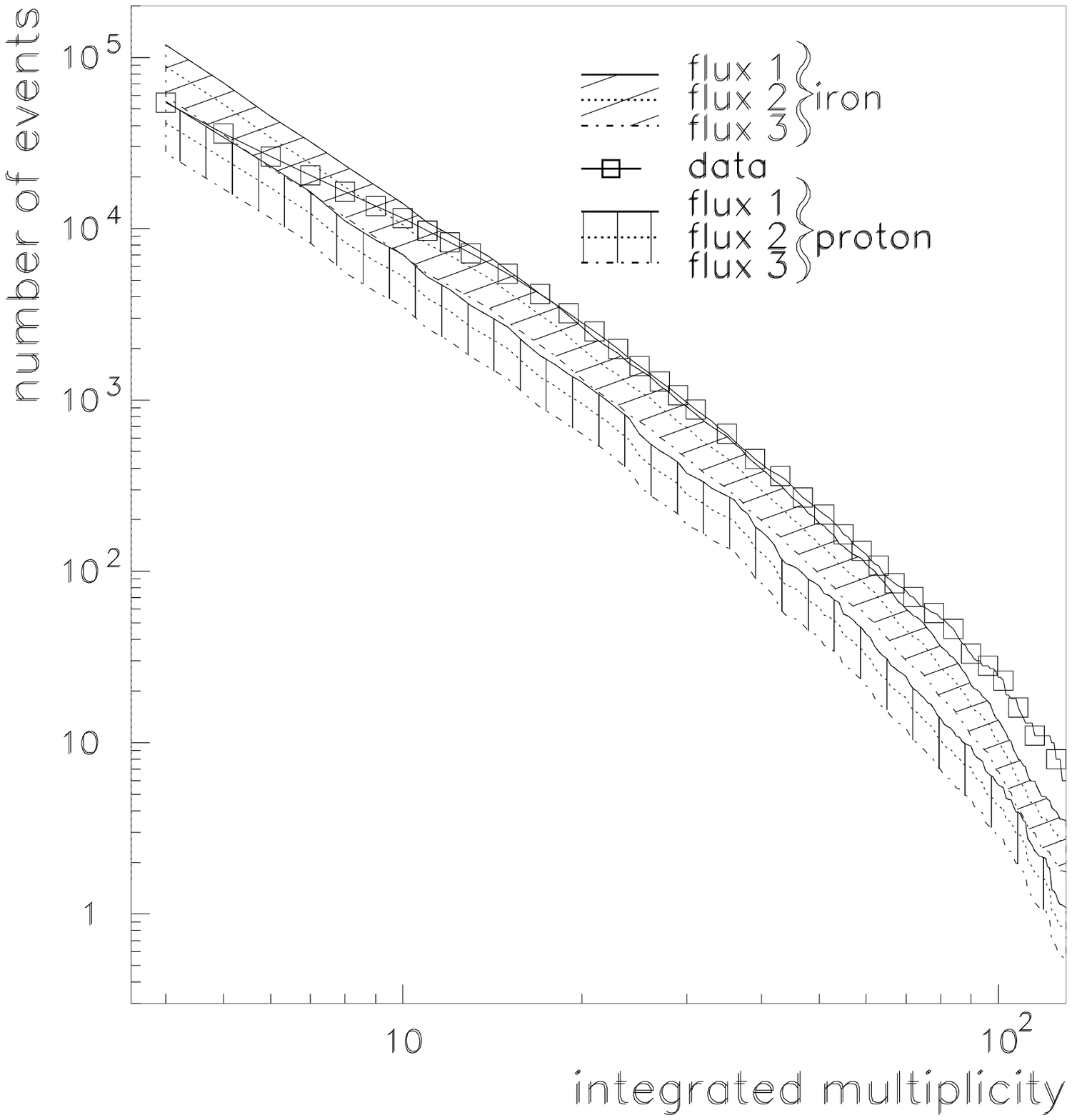,width=7.3cm, height=7.3cm}\\
 a) \hspace{7.5cm} b)
\bf \caption{\rm Integrated multiplicity measured in HAB together with the
  result of the MC simulation of iron and proton induced showers with assumed flux {\it 1} (a) and
  fluxes {\it 1} - {\it 3} (b).}
\label{hacmul}
\end{figure}

The shadowing effect reduces the number of reconstructed tracks when
compared to the number of muons entering the calorimeter. In fact we
measure only a lower limit of the event multiplicity and therefore we plot
the integrated multiplicity distributions where all events with given
multiplicity or higher contribute to the corresponding bin. The
measured distribution is plotted in Fig.~\ref{hacmul}a together with MC
simulations of proton and iron induced showers. 

Taking into account that the composition of
   primary cosmic rays is light at energies $\sim$~$10^{14}$~$\eV$ the
   data should follow the MC prediction for proton primary
   particles at small multiplicities. This behaviour is guaranteed
   only by flux value {\it 1} from Fig.~\ref{asspec}.
 However, this value represents the upper limit of
measured fluxes. Taking into account the spread of flux
{\it 1} - {\it 3}, we obtain for the MC prediction the bands
demonstrated in Fig.~\ref{hacmul}b.

\begin{figure}
\centering
 \epsfig{file=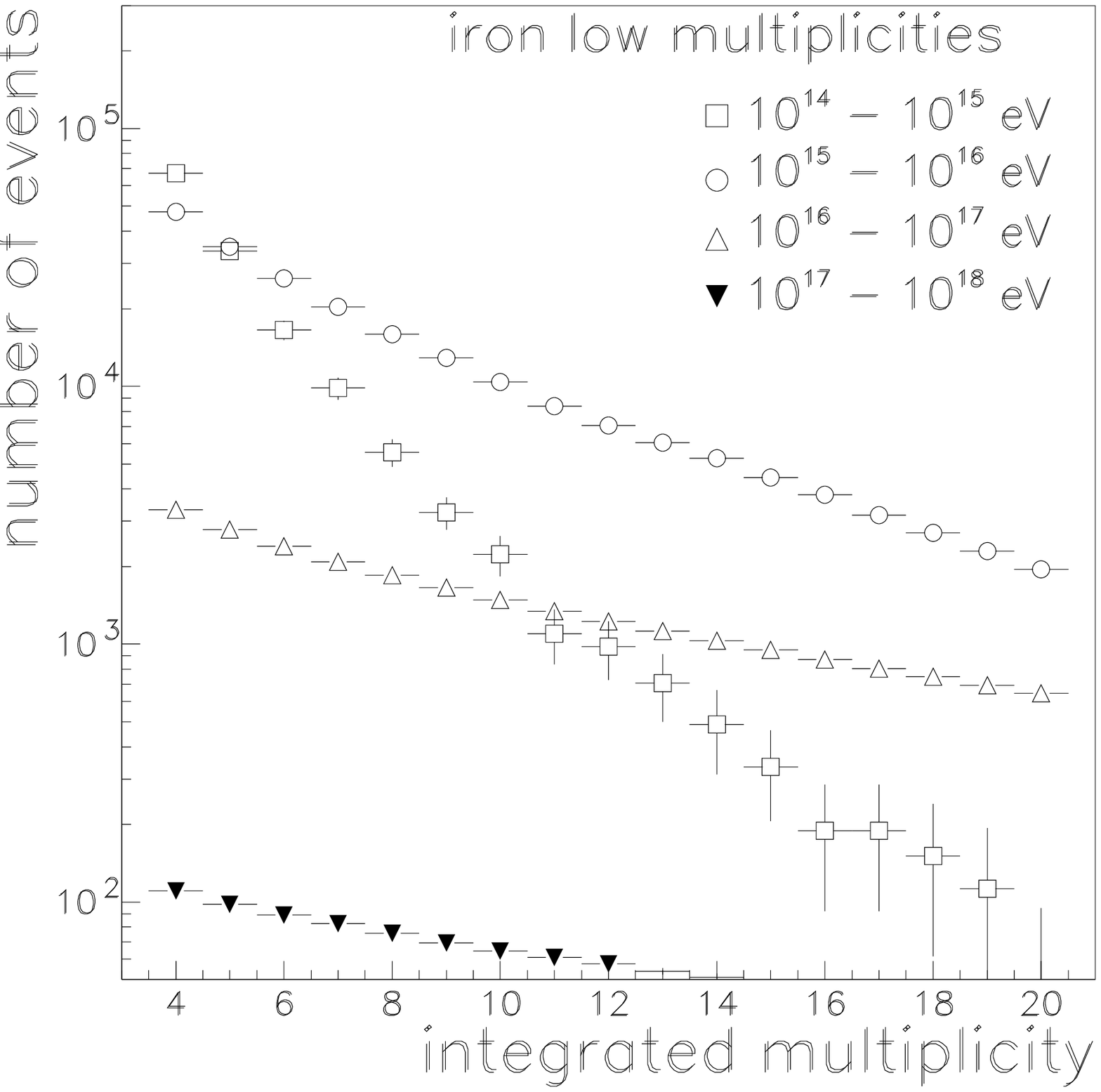,height=7cm}\hspace{1cm}
 \epsfig{file=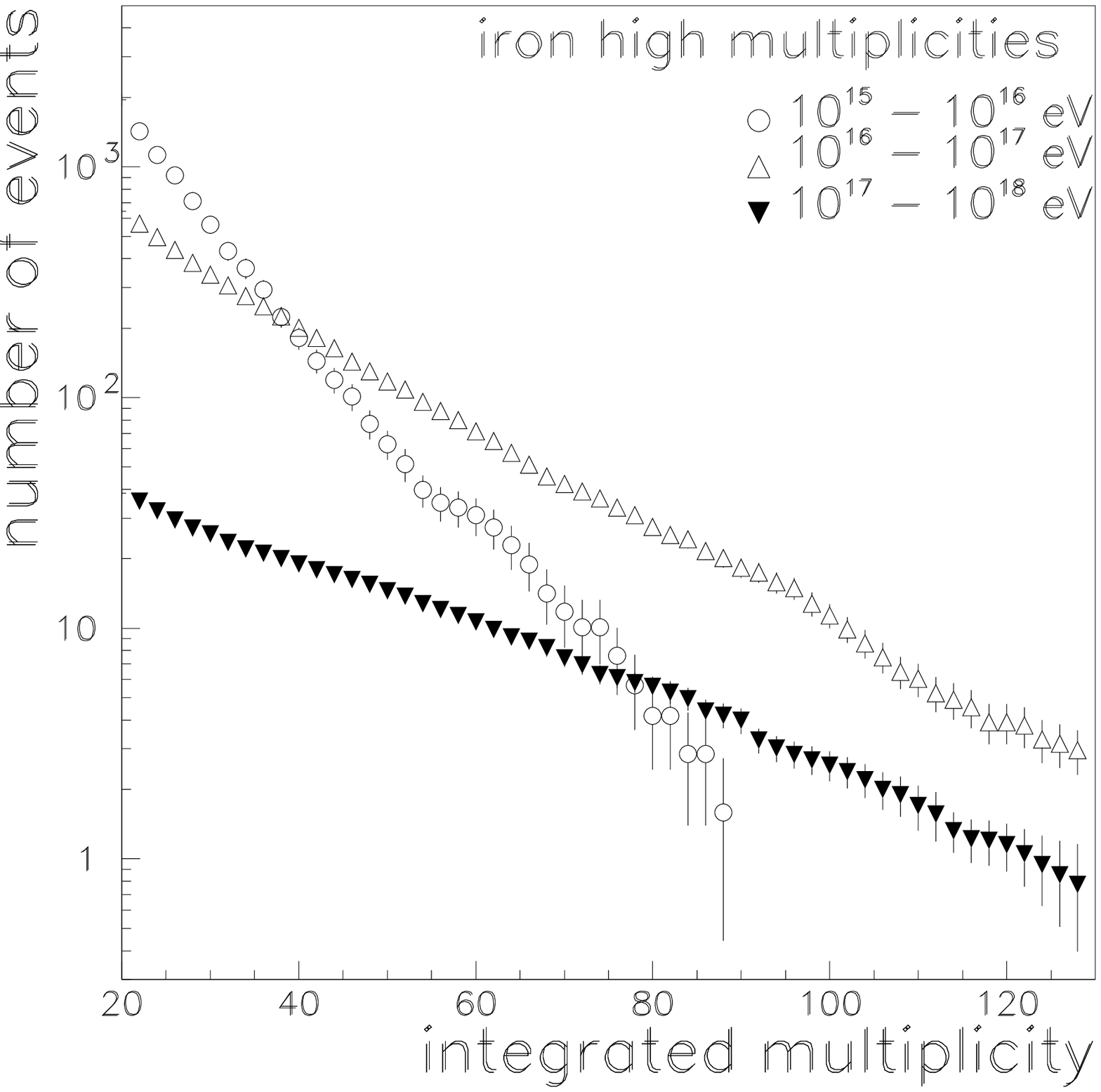,height=7cm}
\bf \caption{ \rm Contributions of different energy intervals to the final
 integral multiplicity distribution. Primary particles are iron nuclei.}
\label{contrfe}
\end{figure}

\begin{figure}
\begin{center}
 \epsfig{file=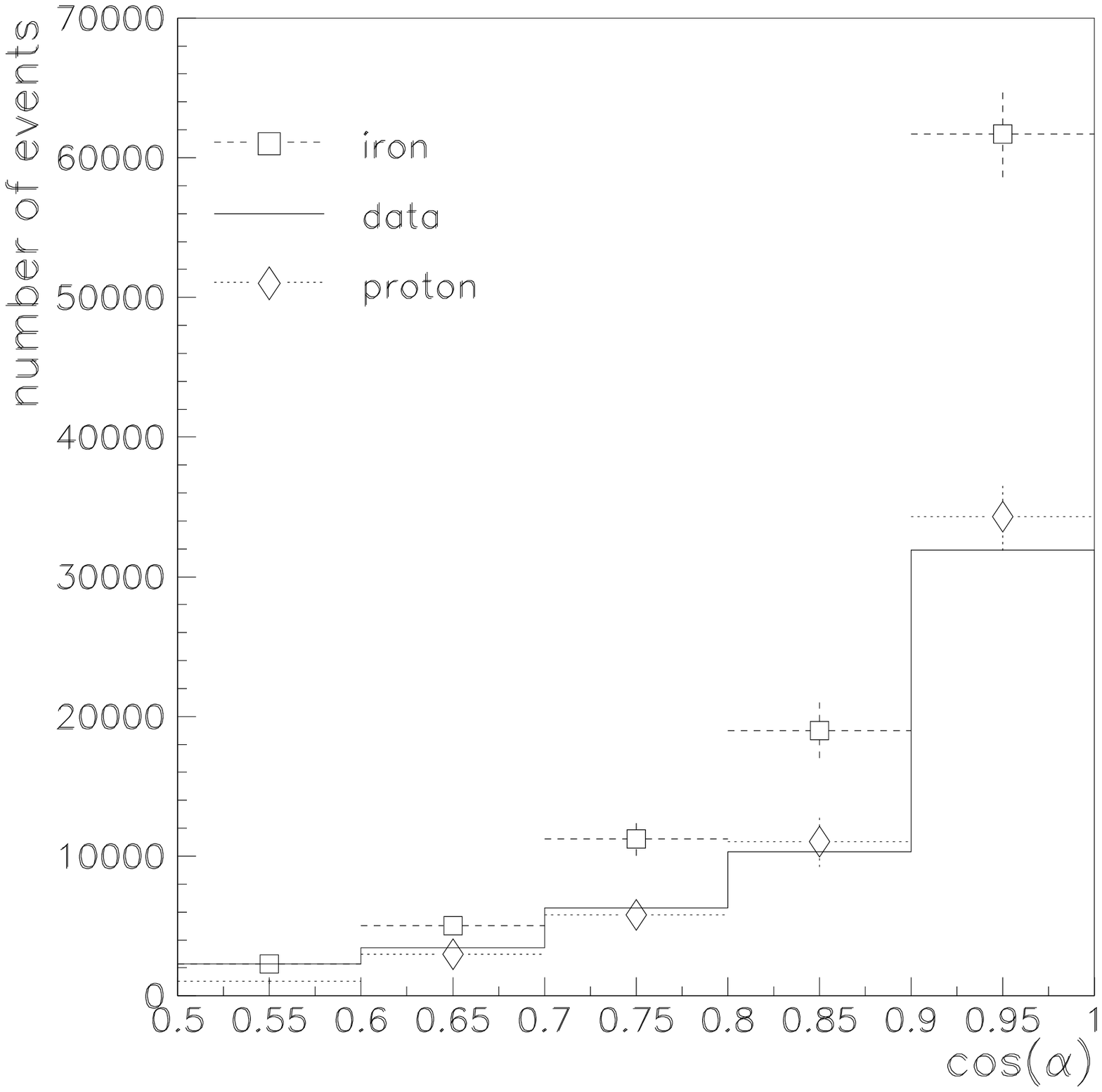,height=7.cm}\hspace{1.5
cm}
 \epsfig{file=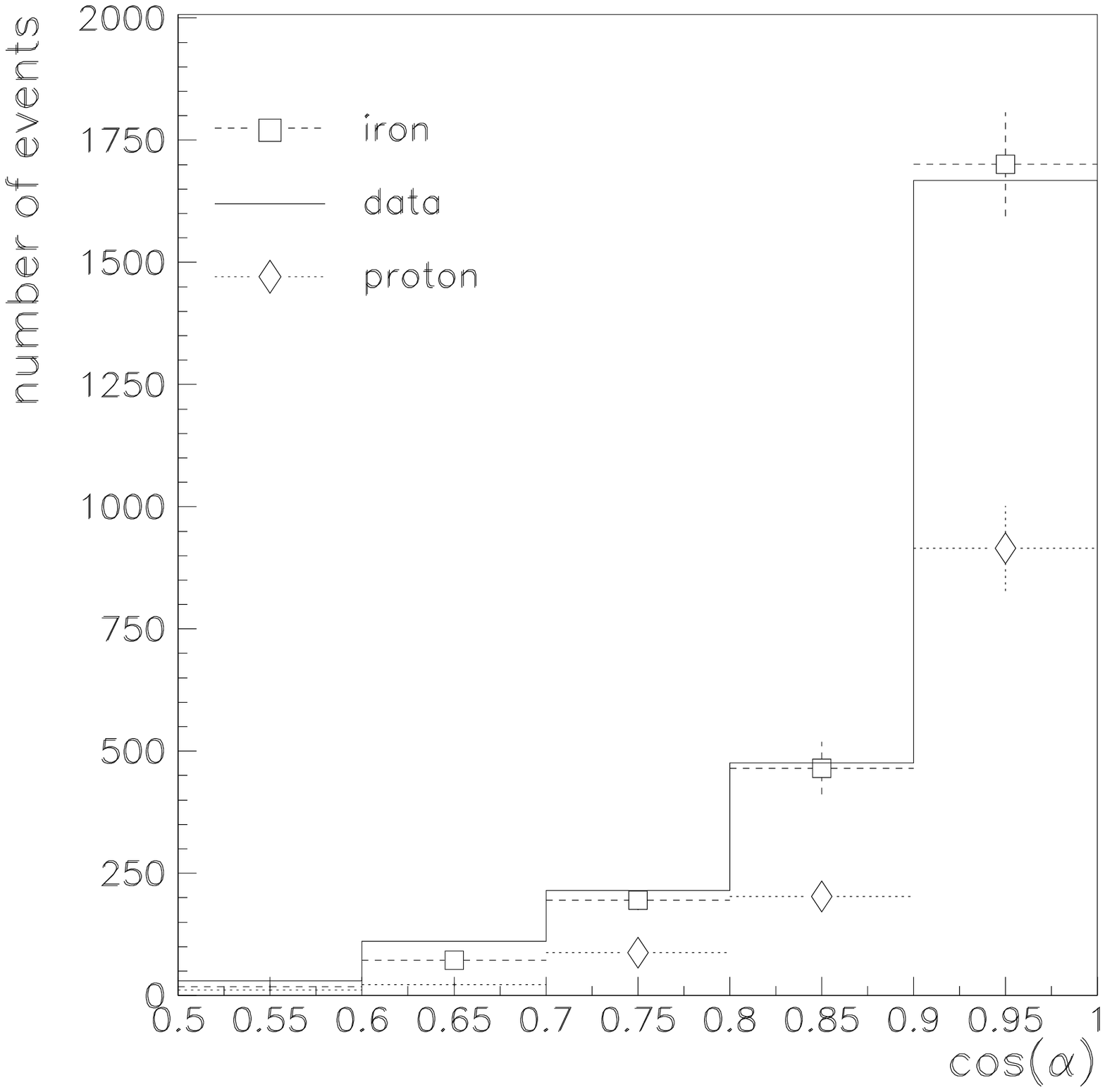,height=7.cm}\\
\end{center}
\hspace{3.7cm} a) \hspace{8.1cm} b)
\begin{center}
\bf \caption{\rm Cosine of the projected angle $\alpha$ at $N_{\mu}
  \geq 4$ (a) and 
  $N_{\mu} \geq 20$ (b) for iron  simulation
  (squares), data (full line) and proton simulation
  (diamonds). Normalisation of MC curves is done according to  flux 1 
      from Fig. \ref{asspec}.}
\label{uhl3}
\end{center}
\end{figure}

Evidently even the highest flux value combined with the assumption of pure
iron primaries is not sufficient to describe the surplus of high
multiplicity events. The excess of events in the region 
$N_{\mu} \geq 80$ is $1.9 \ \sigma$  (based on statistical errors)
for flux {\it 1}; assuming a more
realistic flux value {\it 2}, the discrepancy reaches about $3 \
\sigma$. One is tempted to interpret Fig.~\ref{hacmul} as a convolution
of proton and iron induced showers. However, this would mean that the
primary particles at lower energies would be only protons while at the
higher energies the primaries would be entirely iron nuclei. The
contributions of individual energy bins in the case of iron primaries are
detailed in Fig. \ref{contrfe}. 
 Fig.~\ref{uhl3} shows the distribution of
 projected angle $\alpha$ measured in HAB as compared to MC event samples
 with  $N_\mu \geq 4$ and $N_\mu \geq 20$ respectively. The
lower multiplicity corresponds to the point in Fig.~\ref{hacmul} where
data can be described by proton primaries. The second multiplicity
interval represents the region where 
MC simulation of iron nuclei best describes the data. 

The saturated events appear in the simulation in the same way as in
the data as events with more than 50\% of the tubes hit.
 In the case of primary protons and flux {\it 1} the number of MC
saturated events is $1.1~\pm~0.4$. In the case of iron primaries the total
number of expected saturated events is $3.3~\pm~1.1$ compared to $7$
saturated events in the real data.

\begin{figure}[h!]\centering
 \epsfig{file=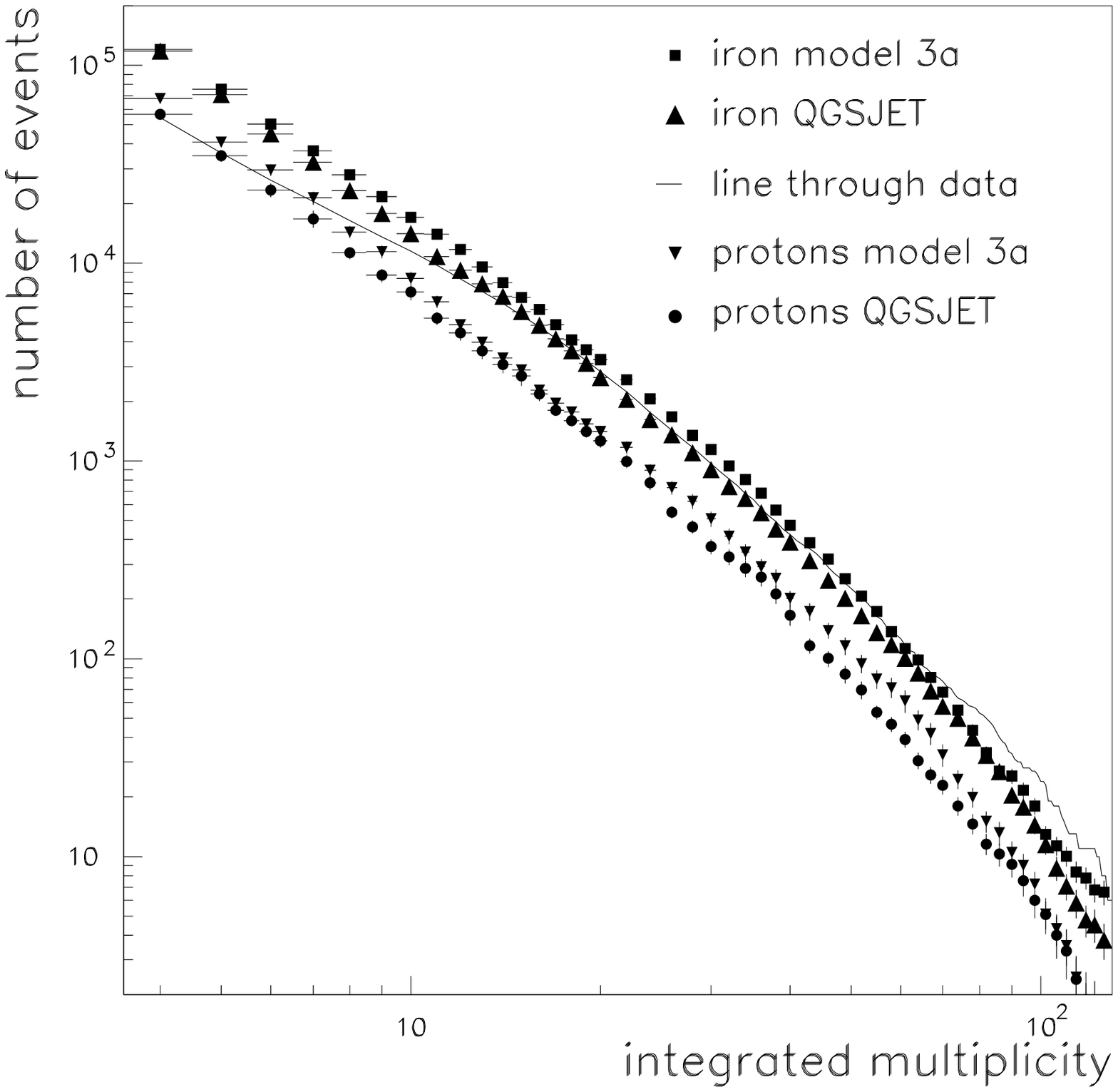,width=7cm, height=7.4cm} \hspace{1.5
cm}
 \epsfig{file=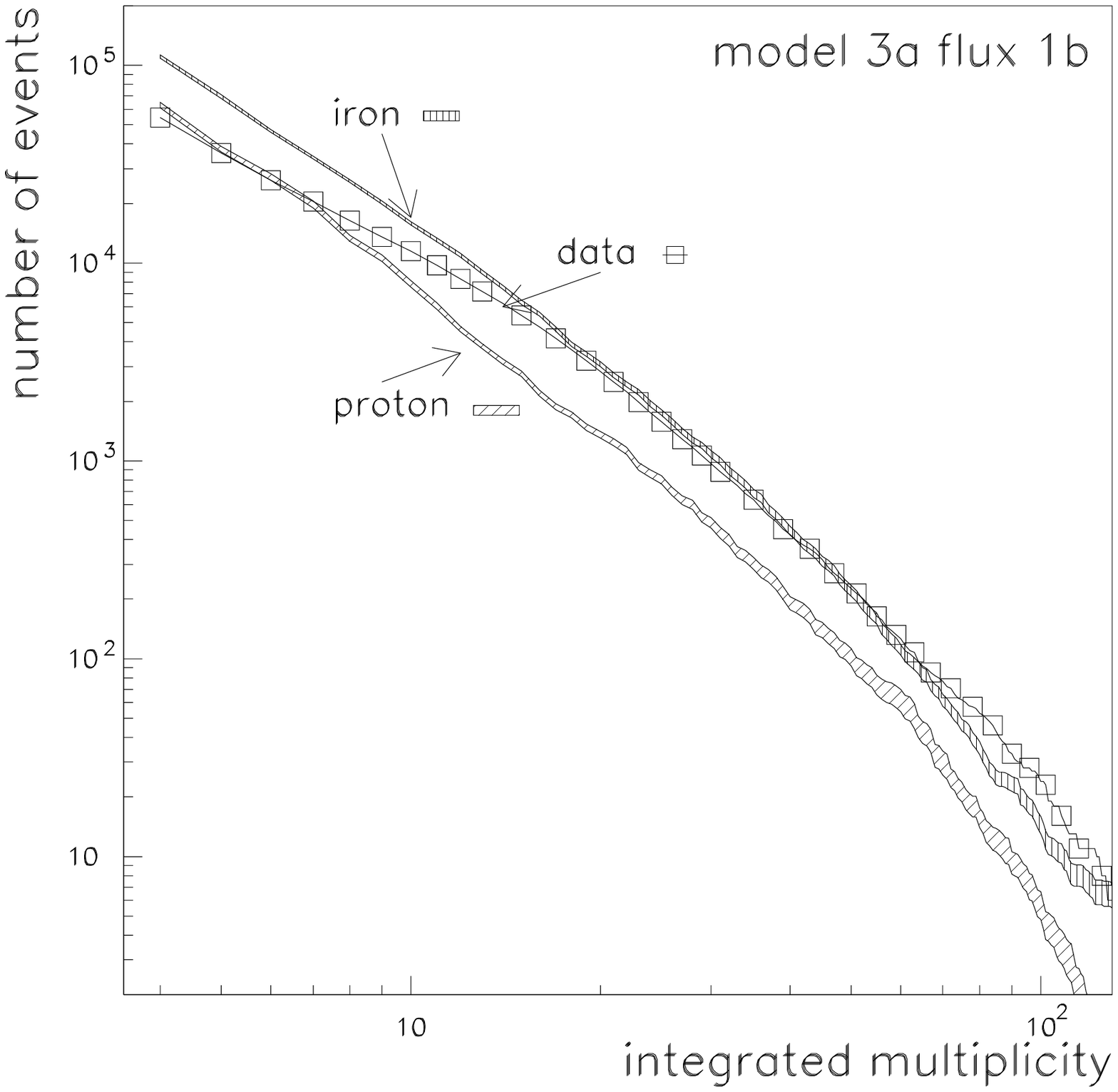,width=7cm, height=7.4cm}

a) \hspace{8cm} b)
\bf \caption{\rm (a) The integral multiplicity distribution for QGSJET and
  modification {\it 3a} compared to data. Flux {\it 1} is assumed. (b)  The integral multiplicity distribution for the modification
  {\it 3a} compared to data. Flux {\it 1b} is assumed.  }
\label{model3apor}
\end{figure}

Although we have tested only the QGSJET model, it is clear that the
use of other models would lead to an even greater discrepancy as
QGSJET predicts higher muon densities close to the shower core than other
models do (e.g. SIBYLL or DPMJET \cite{DPMJET}).  Because of this, it
was suggested 
\cite{diskusehoreng} to test the sensitivity of the produced
multiplicity spectra to QGSJET internal parameters. In
\cite{horandelmod} a set of QGSJET01 parameters is modified; namely
the inelastic cross-section of {\it p-p} ({\it p-N}) is reduced and
the elasticity of the collisions is increased. It is argued that such
modifications can improve consistency between measurements of cosmic
ray composition by experiments based on shower arrays and by
Cerenkov or fluorescence telescopes. Reference \cite{horandelmod}
suggests several possible modifications. In the following we will keep
its notation and denote the tested model as modification {\it 3a}. The
result obtained with the modified QGSJET is compared with the data and
with the original QGSJET01 in Fig.~\ref{model3apor}a.

The model {\it 3a} enlarges the region where the data are between
the predictions for proton and iron induced showers. In the case of unmodified
QGSJET01, the data reach the iron curve at multiplicity $\sim
20$. Using {\it 3a}, the data are consistent with a mixture of light
and heavy components up to a multiplicity $\sim 70$. The slight event
excess in data is still apparent at the highest multiplicities,
however, now with somewhat smaller significance. 

The number of events at low muon multiplicities in the case of proton
primaries and model {\it 3a}  (~Fig. \ref{model3apor}a) is now
larger  than in the 
data. The smaller and more realistic  
flux {\it 1b}   predicts a number of low multiplicity events
consistent with 
 the data as seen from Fig.~\ref{model3apor}b. At high multiplicities
 the  model {\it 3a} predicts of 
 course less events 
 with flux {\it 1b} than with flux assumption {\it 1}. However, the
 prediction of model {\it 3a} with spectrum {\it 1b}  is still above
 the prediction of QGSJET with  flux {\it 1}.

\section{Conclusions}
\label{disc}
The fine granularity hadron calorimeter of the \delphi experiment was
used to  measure multi-muon events originating from cosmic ray
showers. The multiplicity distribution of muon bundles  
cannot be described by current Monte Carlo
models in a satisfactory way. It is difficult to express the
disagreement quantitatively as we have to use flux values measured
elsewhere and also the chemical composition of initial particles is
not well known. However, even the combination of extreme assumptions
of highest measured flux value and pure iron spectrum fails to
describe the abundance of high multiplicity events. 
Similar qualitative conclusions can be drawn from measurements of 
 ALEPH \cite{eggert} and L3+C \cite{henric}, where muon bundles
 (up to multiplicity of about $\sim~30$) were studied in coincidence
 with the ground array signals. 

The tested QGSJET-based model with modified
cross-sections  \cite{horandelmod} performs somewhat better but it
uses a value of the {\it p-p} 
total cross-section at the lowest limit allowed by CDF~\cite{cdf},
E710~\cite{e710} and E811~\cite{e811}
measurements.
Justification of this assumption can be given only by future experiments.
Hadron interactions at energies beyond the reach of accelerators are
not very well known.  Recently, also a more exotic
explanation \cite{strangelets}, based on the assumption of the presence of
strangelets in cosmic rays,  has been suggested to describe
enhanced production of high multiplicity multi-muon events.
 
The main conclusion is that the multi-muon
data from cosmic ray showers detected at intermediate depths are quite
sensitive to the dynamics of initial high energy interactions. In our
case the primary collisions leading to high multiplicity events
($N_{\mu}~>~45$) correspond to interactions at energies equivalent to
about 5 TeV 
in the {\it pp} centre-of-mass system. This energy region has been so far
inaccessible to laboratory measurements. However, even after LHC data
become available,  muon underground measurements have the potential to
reveal some details of interactions in the very forward direction which are
inaccessible to collider experiments. Thus they are important for
the tuning of high energy interaction models 
 which are indispensable for
measurements and energy reconstruction of cosmic rays at even higher
energies of the order $10^{20}$ eV, inaccessible to present and near future
accelerators.

\subsection*{Acknowledgements}
\vskip 3 mm
 We are greatly indebted to our technical collaborators and to the funding agencies for their
support in building and operating the \delphi detector.

We acknowledge in particular the support of \\
Austrian Federal Ministry of Education, Science and Culture,
GZ 616.364/2-III/2a/98, \\
FNRS--FWO, Flanders Institute to encourage scientific and technological
research in the industry (IWT) and Belgian Federal Office for Scientific,
Technical and Cultural affairs (OSTC), Belgium,  \\
FINEP, CNPq, CAPES, FUJB and FAPERJ, Brazil, \\
Ministry of Education of the Czech Republic, project LC527, \\
Academy of Sciences of the Czech Republic, project AV0Z10100502, \\
Grant Agency of the Czech Republic GACR, 202/06/P006, \\
Commission of the European Communities (DG XII), \\
Direction des Sciences de la Mati$\grave{\mbox{\rm e}}$re, CEA, France, \\
Bundesministerium f$\ddot{\mbox{\rm u}}$r Bildung, Wissenschaft, Forschung
und Technologie, Germany,\\
General Secretariat for Research and Technology, Greece, \\
National Science Foundation (NWO) and Foundation for Research on Matter (FOM),
The Netherlands, \\
Norwegian Research Council,  \\
State Committee for Scientific Research, Poland, SPUB-M/CERN/PO3/DZ296/2000,
SPUB-M/CERN/PO3/DZ297/2000, 2P03B 104 19 and 2P03B 69 23(2002-2004)\\
FCT - Funda\c{c}\~ao para a Ci\^encia e Tecnologia, Portugal, \\
Vedecka grantova agentura MS SR, Slovakia, Nr. 95/5195/134, \\
Ministry of Science and Technology of the Republic of Slovenia, \\
CICYT, Spain, AEN99-0950 and AEN99-0761,  \\
The Swedish Research Council,      \\
Particle Physics and Astronomy Research Council, UK, \\
Department of Energy, USA, DE-FG02-01ER41155. \\

\begin{table}[b!]
\begin{center}
\begin{tabular}{|c|c|}
\hline
   & { number of events}\\
\hline
{ $N_{\mu}>3$} &{ $54201$ }\\
{$N_{\mu}\ge 30$ }& { $1065$}\\
{$N_{\mu}\ge 70$} &{ $78$}\\
{ $N_{\mu}\ge 100$ }&{ $24$}\\
\hline
\end{tabular}
\end{center}
\caption{Multiplicities of reconstructed events.}
\label{stat}
\end{table}

\newpage

\newpage

\newpage

\newpage

\newpage

\newpage

\newpage

\newpage

\newpage

\begin{thebibliography}{99}

\bibitem{nexus} H. J. Drescher {\it et al.}, Phys. Rep. {\bf 350} (2001) 93.

\bibitem{qgsjet} N. N. Kalmykov {\it et al.}, 
        Nucl.\ Phys.\ Proc.\ Suppl.\  {\bf 52B} (Issue 3) (1997) 17.

\bibitem{sibyll} R. S. Fletcher {\it et al.},
        Phys.\ Rev.\ D {\bf 50} (1994) 5710.
 
\bibitem{kascademuon} T. Antoni {\it et al.} [KASCADE Collaboration],
  Astropart.\ Phys.\  {\bf 16} (2002) 373. 

\bibitem{casa-mia} K. G. Gibbs  {\it et al.} [CASA Collaboration],
  Nucl.\ Instr.\ 
  Meth.\ A {\bf 264} (1988) 67.

\bibitem{frejus} Ch. Berger {\it et al.} [FREJUS Collaboration],
  Phys. Rev. D {\bf 40} (1989) 2163. \\
 Ch. Berger {\it et al.} [FREJUS Collaboration], Z. Phys. C {\bf 48} 
  (1990) 221. 

\bibitem{KGFmulti} H. Adarkar {\it et al.}, Phys.\ Lett.\ B {\bf 267}
  (1991) 138. 

\bibitem{macro} M. Aglietta {\it et al.} [MACRO and EAS-TOP Collaboration],
  Astropart.\ Phys.\ {\bf 20} (2004) 641.
 
\bibitem{baksan} Yu. M. Andreyev {\it et al.},
         in {\em Proceedings of the 21st Cosmic Ray Conference},
        Adelaide, Australia, 1990, Vol. 9, p. 301.

\bibitem{eggert} V. Avati {\it et al.}, Astropart.\ Phys.\ {\bf 19}
  (2003) 513. 

\bibitem{l3c} O. Adriani {\it et al.}  [L3 Collaboration], Nucl.\
  Instr.\ Meth.\ A {\bf 
    488} (2002) 209. 

\bibitem{Testmodely} 
T.~Antoni {\it et al.}  [KASCADE Collaboration],
J.\ Phys.\ G {\bf 25} (1999) 2161.

\bibitem{delphidet} P. Aarnio {\it et al.} [DELPHI Collaboration],
  {Nucl. Instr. Meth.}  A     {\bf 303} (1991) 233.

\bibitem{delphidet2} P. Abreu {\it et al.} [DELPHI Collaboration],
  Nucl. Instr. Meth.  
         A {\bf 378} (1996) 57.
\bibitem{aj1}
 I. Azinenko {\it et al.},
IEEE Trans. on Nucl. Science {\bf NS-42}   No.4  (1995)  485.
\bibitem{aj2}
 I. Azinenko {\it et al.},
 IEEE Trans. on Nucl. Science {\bf NS-43} No.3. (1996)  1751.


\bibitem{diser} P. Travnicek, PhD. thesis, Charles University, Prague 2004,
         CERN-THESIS-2006-032 also available at: www-hep2.fzu.cz/$\sim$travnick/thesis.ps.gz.


\bibitem{ECTANA} J. Ridky, V. Vrba, J. Chudoba,  DELPHI NOTE 99-181
  TRACK 96, (1999). 

\bibitem{CORSIKA}  D. Heck  {\it et al.},
         FZKA-6019, (1998), Forschungszentrum Karlsruhe.

\bibitem{GEANT} R. Brun {\it et al.}, GEANT3, {\it Report CERN
    DD/EE/84-1} (1984), CERN, Geneva. 

\bibitem{delsim} DELPHI Collaboration, DELPHI NOTE: 89-67 PROG 142, (1989).

\bibitem{pdg} K. Hagiwara {\it et al.} [Particle Data Group],
         Phys.  Lett. B {\bf 592}  (2004) 1.


\bibitem{have} M. A. Lawrence  {\it et al.}, J. Phys. G 
{\bf 17}   (1991) 733.

\bibitem{DPMJET} J.~Ranft, Phys.\ Rev.\ D {\bf 51} (1995) 64.

\bibitem{diskusehoreng} Ralph Engel and J\"org H\"{o}randel, private
  communication. 

\bibitem{horandelmod} J. R. H\"orandel, J.\ Phys.\ G {\bf 29} (2003)
  2439.

\bibitem{henric}        H. Wilkens  [L3 Collaboration],
  {\em Proceedings of the 28th Cosmic Ray Conference}, Tsukuba 
        2003, Cosmic Ray 1131-1134.

\bibitem{cdf} F. Abe {\it et al.} [CDF Collaboration], Phys.\ Rev.\ D {\bf 50} (1994) 5550.

\bibitem{e710} N. A. Amos {\it et al.} [E710 Collaboration], Phys.\ Lett.\ B {\bf 243} (1990) 158.

\bibitem{e811} C. Avila {\it et al.} [E811 Collaboration], Phys.\ Lett.\ B {\bf 445} (1999) 419.

\bibitem{strangelets} M. Rybczynski, Z. Wlodarczyk and G.~Wilk, Acta
  Phys.\ Polon.\ B {\bf 33} (2002) 277. 



\end{thebibliography}
\end{document}